# Unlocking the Inaccessible Energy Density of Sodium Vanadium Fluorophosphate Electrode Materials by Transition Metal Mixing

Stephanie C. C. van der Lubbe[1#], Ziliang Wang[1#], Damien K. J. Lee[2], and Pieremanuele Canepa[1,2*]

[1]Department of Materials Science and Engineering, National University of Singapore, 9 Engineering Drive 1, 117575, Singapore

[2]Department of Chemical and Biomolecular Engineering, National University of Singapore, 4 Engineering Drive 4, 117585, Singapore

[#]These authors contributed equally to this work

*pcanepa@nus.edu.sg

## Abstract

Sodium (Na) vanadium (V) fluorophosphate $Na_xV_2(PO_4)_2F_3$ (NVPF) is a highly attractive intercalation electrode material due to its high operation voltage, large capacity, and long cycle life. However, several practical issues limit the full utilization of NVPF's energy density: 1) the high voltage plateau associated with extracting the "third" Na ion in the reaction $N_1$VPF –> VPF (~4.9 V vs Na/$Na^+$) appears above the electrochemical stability window of most practical electrolytes (~4.5 V vs Na/$Na^+$); 2) a sudden drop in Na-ion diffusivity is observed near composition $Na_1V_2(PO_4)_2F_3$. Therefore, it is important to investigate the potential substitution of V by other transition metals in NVPF derivatives, which can practically access the extraction of the third Na-ion. In this work, we investigate the partial substitution of V with molybdenum (Mo), niobium (Nb), or tungsten (W) in NVPF to improve its energy density. Using first-principles calculations, we examine the structural and electrochemical behaviors of $Na_xV_{2-y}Mo_y(PO_4)_2F_3$, $Na_xV_{2-y}Nb_y(PO_4)_2F_3$, and $Na_xW_2(PO_4)_2F_3$ across the whole Na composition region of $0 \leq x \leq 4$, and at various transition metal substitution levels, namely, y = 0.5, 1.0, 1.5, 2.0 for Mo, and y = 1.0, 2.0 for Nb. We find that partial substitution of 50% V by Mo in NVPF reduces the voltage plateau for extracting the third Na ion by 0.6 Volts, which enables further Na extraction from $Na_1VMo(PO_4)_2F_3$ and increases the theoretical gravimetric capacity from ~128 to ~174 mAh/g. Analysis of the migration barriers for Na-ions in $Na_xVMo(PO_4)_2F_3$ unveils improved kinetic properties over NVPF. The proposed $Na_xVMo(PO_4)_2F_3$ material provides an optimal gravimetric energy density of ~577.3 Wh/kg versus ~507 Wh/kg for the pristine NVPF, which amounts to an increase of ~13.9%.

## Introduction

With a larger natural abundance of sodium (Na) and lower raw material costs, sodium-ion batteries offer a promising alternative to lithium-ion batteries.[1–3] However, the higher reduction potential for Na (–2.71 Volts versus standard hydrogen electrode (SHE)) than for Li (–3.04 Volts versus SHE), larger ionic size, and greater mass of Na contribute to a lower theoretical energy density.[1,2] Therefore, developing energy-dense electrode materials based on sustainable constituents is a key challenge.

Polyanion-type materials are widely investigated as positive electrode materials for Na-ion batteries due to their high voltages arising from inductive effects.[4–8] The sodium vanadium fluorophosphate $Na_xV_2(PO_4)_2F_3$ (NVPF) is one of the most attractive positive electrodes for Na-ion batteries due to its high operation voltage (~3.95 V vs. $Na/Na^+$ on average),[9] large capacity (~128 mAh/g),[10] and long cycle life of 100% specific capacity retention after 10,000 cycles.[11–14] In practice, two Na-ions can be reversibly extracted/inserted between $Na_3V_2(PO_4)_2F_3$ ($N_3VPF$) and $Na_1V_2(PO_4)_2F_3$ ($N_1VPF$), giving a gravimetric energy density of ~507 Wh/kg.[10,15–17]

The potential extraction of three Na-ions from $N_3VPF$ to $V_2(PO_4)_2F_3$ (VPF) (instead of $N_1VPF$) would increase the theoretical gravimetric capacity from 128 mAh/g to ~198 mAh/g. However, extraction of the third $Na^+$ from $N_1VPF$ activates the $V^{IV}/V^{V}$ redox couple at a voltage plateau of ~4.9 V vs. $Na/Na^+$.[18] This voltage appears above the electrochemical stability window (ESW) for most sodium electrolytes,[19] rendering the third Na-ion electrochemically inaccessible. Furthermore, NVPF exhibits a rapid decline in $Na^+$ diffusivity as the Na content (per formula unit, f.u.) approaches 1, which has been attributed to additional energies required to create vacancies and to disrupt strong Na-vacancy orderings at $N_1VPF$.[17,20,21] Recently, the group of Tarascon has demonstrated that the reversible extraction of the third Na-ion in NVPF is possible via the formation of a disordered tetragonal NVPF phase (space group *I4/mmm*). This disordered structure can reversibly uptake three Na-ions, but at a lower potential than NVPF in the original ordered state, resulting in a 14% improvement in energy density.[10] Achieving reversible three $Na^+$ cycling with further optimization of the energy density in NVPF remains an open challenge.

Here, we demonstrate that the energy density of the NVPF positive electrode can be optimized via transition metal (TM) substitution, which replaces the high-potential $V^{IV}/V^{V}$ at $0 \leq x \leq 1$ of NVPF by alternative TM redox couples and enables the reversible cycling of three Na-ions. Using first-principles calculations, we have replaced the vanadium with niobium (Nb), molybdenum (Mo), and tungsten (W) at different substitution levels, forming $Na_xV_{2-y}M_y(PO_4)_2F_3$ (M = Nb, Mo, or W, and $0 \leq x \leq 4$). We have studied the structural, electrochemical, and kinetic properties of these novel $Na_xV_{2-y}M_y(PO_4)_2F_3$ compositions upon Na-ion intercalation. Nb and Mo were selected based on a high-throughput *ab initio* study by Hautier et al., which demonstrated the average voltage of the $Nb^{IV}/Nb^{V}$ and $Mo^{IV}/Mo^{V}$ redox couples to be lower than $V^{IV}/V^{V}$ by ~2.0 and ~0.6 V, respectively.[22] W was included because it is isoelectronic to Mo, and has recently been used as a dopant for enhancing the electrochemical performance of the NVPF cathode ($Na_3V_{1.96}W_{0.04}(PO_4)_2F_3$) material.[23]

By replacing 50% of V in NVPF with Mo, as in $Na_xVMo(PO_4)_2F_3$ (NVMoPF), we find that the voltage plateau associated with the extraction of the third Na-ion from $Na_1VMo(PO_4)_2F_3$ ($N_1$VMoPF) is reduced from 4.53 to 3.93 V, which is practical for current nonaqueous electrolytes.[19] The decreased voltage may enable the reversible extraction of three Na-ions and lead to an increase in capacity from 128 mAh/g for NVPF to 173.7 mAh/g for NVMoPF.

Further in-depth studies into the kinetic properties of NVMoPF reveal the favorable mechanism of extracting the third Na-ion from $N_1$VMoPF due to superior $Na^+$ mobility, making NVMoPF a promising candidate for increasing the reversible capacity in comparison to NVPF. The increased capacity contributes to a boost of gravimetric energy density from ~507 Wh/kg for NVPF ($1 \leq x \leq 3$)[15] to ~577.3 Wh/kg for NVMoPF ($0 \leq x \leq 3$), corresponding to a rise of ~13.9%.

## Results

### Structural Features of the $Na_xV_2(PO_4)_2F_3$ Skeleton

Here, chemical formulae are abbreviated to the first letter of their chemical group throughout this whole manuscript. For example, $Na_xMo_2(PO_4)_2F_3$ is abbreviated as NMoPF, and $Na_xVMo(PO_4)_2F_3$ is abbreviated as NVMoPF. When necessary, the sodium composition x is specified by the subscript, e.g. $Na_1VMo(PO_4)_2F_3$ is written as $N_1$VMoPF. The Na-ion that is extracted from $Na_1$ to $Na_0$ will be referred to as the third $Na^+$.

The host structure of NVPF is composed of $V_2O_8F_3$ bioctahedral units corner-shared by $PO_4$ tetrahedra, resulting in a robust 3-D framework (see **Figure 1a**) and several Na-ion “interstitial” sites.[15,24,25] As shown in **Figure 1**, $PO_4$ polyanionic groups separate the two-dimensional Na layers, which lie in the *ab*-plane of the structure.

The pristine $N_3$VPF phase comprises three distinct sodium positions that form a circle around fluorine centers,[9,13,15,24] as shown in **Figure 1b**. The sodium ions can occupy the [010], [100], and [110] sites, which are indexed as Na(1) (4c), Na(2) (8f), and Na(3) (8f), respectively.

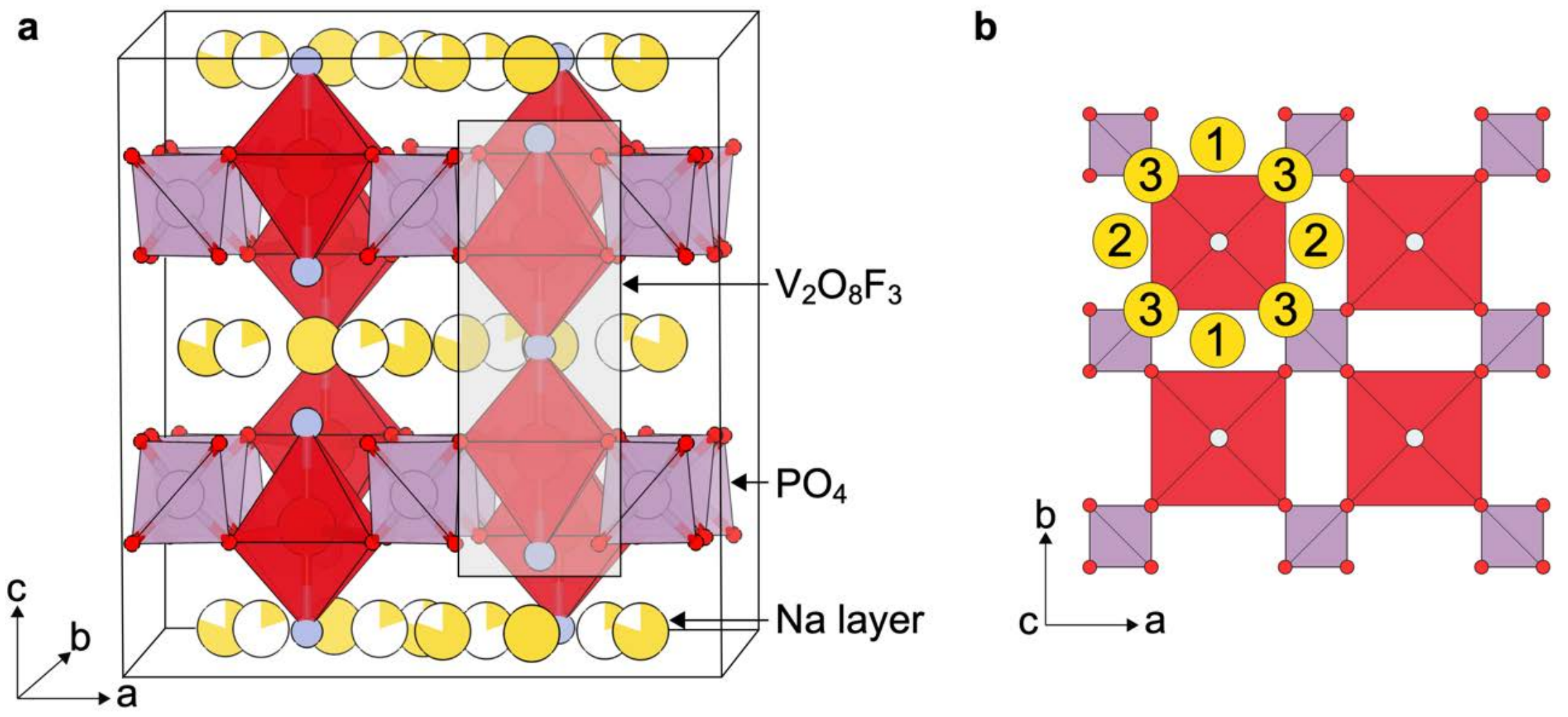

**Figure 1.** The crystal structure of the pristine $Na_3V_2(PO_4)_2F_3$ phase in the orthorhombic *Amam* space group is shown in panel **a**. The $V_2O_8F_3$ bioctahedral unit (red) contains three fluorine atoms (gray) that are aligned along the [001] direction. The bioctahedral units are corner-share with $PO_4^{3-}$ tetrahedra (purple) that are aligned in parallel within the (001) plane, creating two-dimensional Na layers (yellow) in the *ab*-plane of the structure. Panel **b** shows the top view of the *ab*-plane for $Na_xV_2(PO_4)_2F_3$ and the available Na sites and their indexing. Three distinct sodium sites with their Wyckoff labels, i.e., Na(1) (4c), Na(2) (8f), and Na(3) (8f) sites (yellow circles) are indexed with 1, 2, and 3, respectively.

The Na(1) sites are fully occupied, whereas Na(2) and Na(3) sites are ~80% and ~20% occupied, respectively, for $N_3$VPF at room temperature. [15,24,26] At the other Na contents in $N_x$VPF, specific Na-vacancy ordering or disordering will occur among the three Na sites (**Figure 1b**), and thus lead to different symmetries.[15,25] Here, we will extend the same indexing convention of Na positions to NVPF systems where V is (partly) replaced by Mo.

### Phase Stabilities and Voltage Curve of Mo and Nb Substituted $Na_xV_2(PO_4)_2F_3$

Using density functional theory (DFT, see methods section), we computed the energy of mixing (i.e., the formation energy $E_f$) of Na of a substantial number of Na-vacancy orderings for $N_xVPF$ (x = 0 ~ 4 in a step size of $increment$ = 0.5), $N_xVMoPF$ (x = 0 ~ 4 with increments $\Delta x$ = 1) and $N_xMoPF$ (x = 0 ~ 4 with $\Delta x$ = 1), which amounted to 202, 243 and 99 unique DFT-relaxations (atom coordinates + model cell shape + model cell volume) for NVPF, NVMoPF, and NMoPF, respectively. The computed energies of mixing are shown in **Figure 2a-c** as the Na concentration x is varied. **Figure 2d-f** shows the corresponding intercalation voltages vs. $Na/Na^+$ as a function of Na concentration for the parent NVPF and Mo-substituted NVPF systems, namely NVMoPF and NMoPF.

From the analysis of **Figure 2a-c**, we identified the phase diagrams at 0 K –the convex hulls– by the lower envelopes formed by the most stable Na-vacancy orderings in each tie-line considered. Structures with Na-vacancy orderings above the convex hull are metastable or unstable and likely to decompose into the nearest ground state compositions. The ground state structures representing Na-vacancy orderings for NVPF, NVMoPF, and NMoPF were identified at specific Na content x = 0, 1, 2, 3, 4. These ground-state structures feature in the voltage curves of **Figure 2d-f**.

The green shading domains in **Figure 2d-f** indicate the practical Na composition range in which Na-ions are predicted to be accessible for reversible cycling. We consider the Na composition in the range of $0 \leq x \leq 3$ to be accessible if their voltage plateau is below 4.5 V vs. $Na/Na^+$, which is a common threshold for existing nonaqueous electrolyte stability.[13,21,27] Note that the exact onset potential for electrolyte decomposition depends on the type of salt(s) and solvent(s).[27] For example, Ponrouch et al. reported an average value of approximately 4.6 V vs. $Na/Na^+$ for $NaClO_4$ in 10 different organic solvents in the range of 3.7 to 5.2 V vs. $Na/Na^+$.[27] The Na-ion associated with the reversible insertion step in the composition range of 3

≤ x ≤ 4 is considered inaccessible due to 1) the limited concentration of vacancies when approaching the fully discharged $Na_4$ composition, which prevents $Na^+$ mobility,[17] and 2) the low voltage plateaus (~0 V) that fall below the electrolyte stability window.[26,28]

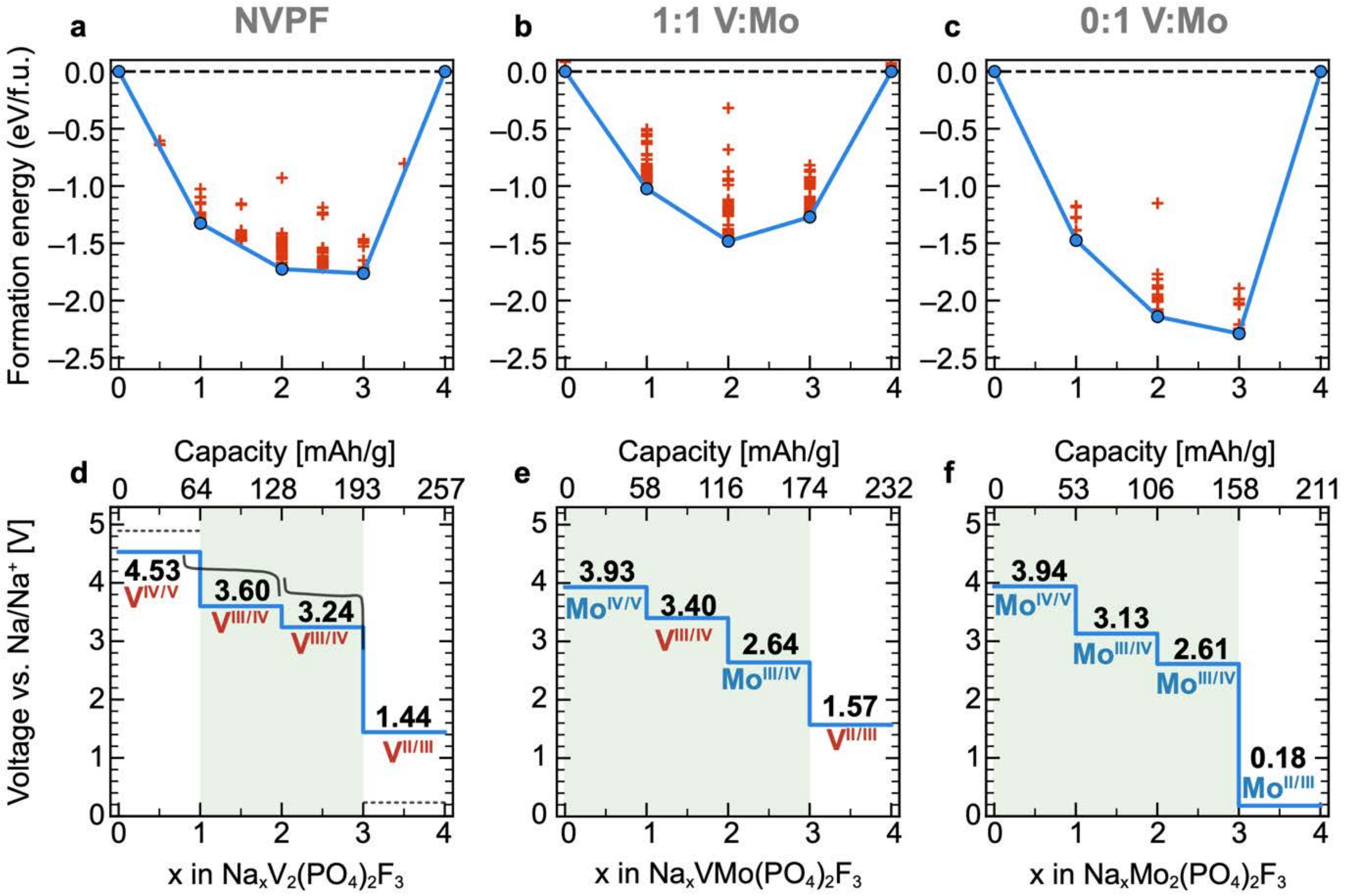

**Figure 2.** Panels **a** to **c** show the computed mixing energies (approximated by formation energy) with GGA+*U* as a function of Na concentration x for $N_xVPF$ (**a**), $N_xVMoPF$ (**b**), and $N_xMoPF$ (**c**). Unstable Na-vacancy orderings are shown by the red crosses, whereas ground state orderings (blue dots) form the phase diagram at 0 K — the convex hull. Panels **d** to **f** depict the corresponding intercalation voltages vs. $Na/Na^+$ as a function of Na concentration x for $N_xVPF$ (**d**), $N_xVMoPF$ (**e**), and $N_xMoPF$ (**f**). The computed voltage values and the activated redox couples for each step are signified. The experimental voltage of NVPF is superimposed with solid black lines in (**d**), depicting the (de)intercalation step of $Na_1$-$Na_2$ and $Na_2$-$Na_3$[15] (reproduced and adapted with permission from Ref. 15. Copyright 2015 American Chemical Society). For the impractical high-voltage step at $Na_0$-$Na_1$[18] (reproduced and adapted with permission from Ref. 18. Copyright 2014 American Chemical Society), and low-voltage step at $Na_3$-$Na_4$ for NVPF [26,28] (reproduced and adapted with permission under a Creative Commons 4 from Ref. 26. Copyright 2016 Springer Nature. Reproduced and adapted with permission from Ref. 28 2017 John Wiley and Sons) only the average voltages are schematically shown with dashed black lines for visualization (**d**). All capacities are normalized using the mass of each fluorophosphate structure at x = 3. The Na composition region with accessible capacity is indicated by green shades. Voltage curves and convex hulls of the $N_xV_{1.5}Mo_{0.5}PF$ and $N_xV_{0.5}Mo_{1.5}PF$ systems are in **Supporting Figure S1**.

We start by comparing the computed voltage profiles to the available experimental voltages for NVPF.[18,19,29] Our results show two voltage plateaus at ~3.24 and ~3.60 V vs. Na/$Na^+$, respectively, which are associated with the reversible $Na^+$ extraction from $N_3$VPF to $N_1$VPF, and one voltage plateau at ~4.53 V vs. Na/$Na^+$ that is associated with the extraction of the third $Na^+$ from $N_1$VPF to VPF, which is not reversible in experiments (**Figure 2d**).[15,18] These plateaus were observed in previous experiments at ~3.6, ~4.2, and ~4.9 V vs. Na/$Na^+$, respectively.[18,19,29] The discrepancy between the theoretical and experimental value of voltage has previously been attributed to the Hubbard-*U* correction of 3.1 eV for V centers, which causes a constant voltage underestimation of approximately ~0.4 V.[21] Importantly, the intercalation trends predicted in our calculations are in line with the experimental results, which underlines the reliability of our method for predicting the electrochemical properties.

Following the reaction of $N_1$VPF –> VPF, the voltage associated with the third $Na^+$ extraction from $N_1$VPF appears above the electrochemical stability window of 4.5 Volts of the most common nonaqueous electrolytes for Na-ion batteries,[13,21,27] thus making this $Na^+$ inaccessible.[19] As shown in **Figure 8** and **Supporting Table S1**, extracting two Na-ions at the predicted voltages of ~3.60 and ~3.24 V vs. Na/$Na^+$ results in a theoretical energy density of ~439.1 Wh/kg for NVPF (~507 Wh/kg as observed in experiments).[15]

The replacement of all V with Mo (giving NMoPF) decreases the values of the voltage plateaus by ~1.26, ~0.63, ~0.47, and ~0.59 V in comparison to NVPF upon $Na^+$ extraction from the hypothetical fully discharged $N_4$MoPF phase to the empty MoPF structure, respectively. From our calculations, we observe that the third $Na^+$ extraction from $N_1$MoPF to MoPF activates the $Mo^{IV}/Mo^{V}$ redox couple at a voltage of

~3.94 V vs. Na/$Na^+$. This is clearly below the electrolyte stability window of the most common electrolytes.[27]

When substituting 50% of Mo for V, forming NVMoPF, the $Na^+$ extraction from $N_3$VMoPF to the empty VMoPF occurs through the activation of the redox couples of $Mo^{III}/Mo^{IV}$, $V^{III}/V^{IV}$, and $Mo^{IV}/Mo^{V}$ (see **Figure 2e**). The extraction of the third $Na^+$ from $N_1$VMoPF shows a voltage of ~3.93 V vs. Na/$Na^+$, which is similar in magnitude to the same redox process in the all-Mo-containing electrode (~3.94 vs. Na/$Na^+$). The theoretical energy density of NVMoPF (x = 0 ~ 3) is 577.3 Wh/kg, which is ~13.9% higher than NVPF (considering the third $Na^+$ extraction an irreversible process for NVPF, see **Supporting Table S1**).

We also investigated the structural and electrochemical properties of $Na^+$ extraction from the NNbPF, NVNbPF, and NWPF systems, whose voltage curves are shown in **Figure 3**. The corresponding mixing energy profiles are given in **Supporting Figure S2**.

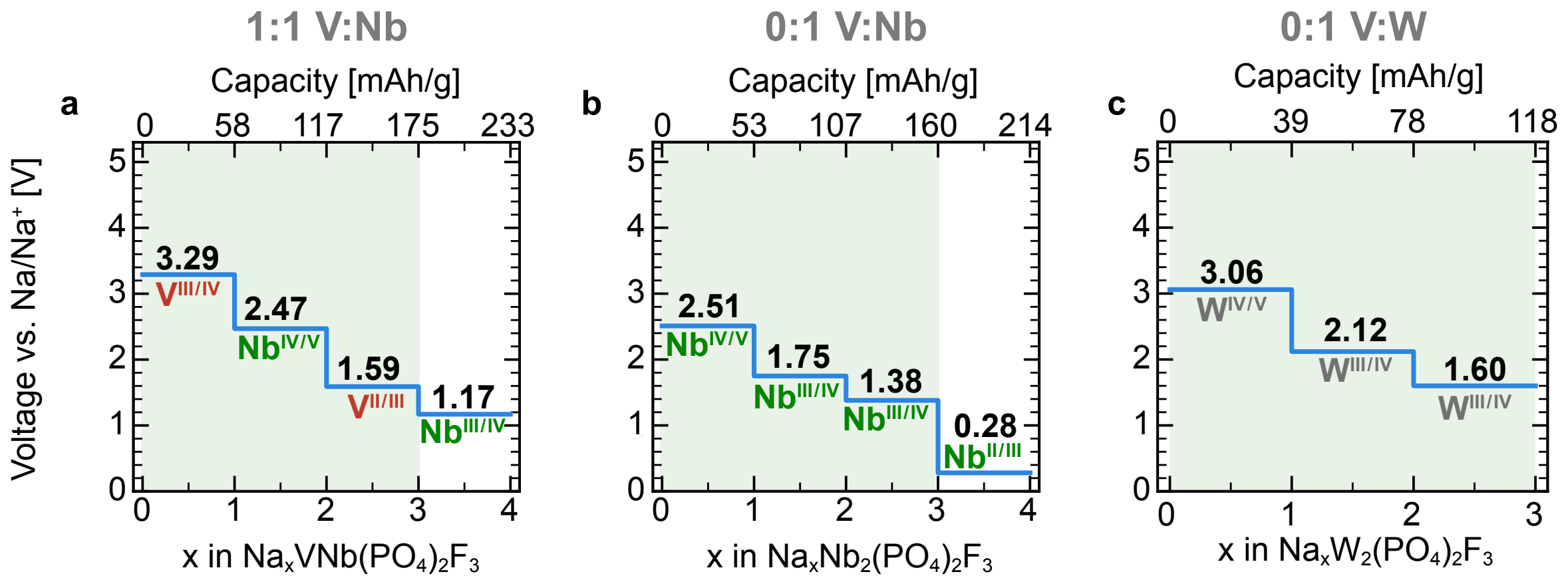

**Figure 3.** The DFT+*U* computed intercalation voltages vs. Na/$Na^+$ as a function of Na concentration x for **(a)** $N_x$VNbPF, **(b)**, $N_x$NbPF and **(c)** $N_x$WPF. The voltage profile of $Na_3$-$Na_4$ in NWPF is omitted because the electronic state of $N_4$WPF is not correctly described by the GGA+*U* framework, leading to unphysical oxidation states and thus an inaccurate prediction of the electronic structure. All the gravimetric capacities are calculated using the masses of composition x = 3. The active redox couples are indicated in red for V, green for Nb, and silver for W. The accessible capacity is indicated by green shades.

On average, the voltage plateaus for the Nb-containing system NNbPF are nearly 2 V lower from x=0 to x=3, and over 1 V lower from x=3 to x=4 compared to NVPF. The extraction of the third Na-ion from x=1 is predicted to occur at ~2.51 V vs. $Na/Na^+$ via $Nb^{IV}/Nb^{V}$. Assuming a potential extraction of three Na-ions, NNbPF could provide almost a 20% increased capacity (160 mAh/g) compared to NVPF (128 mAh/g). However, the higher capacity of NNbPF cannot compensate for the substantially lower intercalation voltages of NNbPF, which reduces the gravimetric energy density for NNbPF (~301 Wh/kg) by 40.6% compared to NVPF.

Surprisingly, in the mixed Nb/V fluorophosphate (i.e., NVNbPF), the extraction of the third $Na^+$ from $N_1$VNbPF to VNbPF activates the $V^{III}/V^{IV}$ redox couple at ~3.29 V (**Figure 3a**), which has a much lower voltage than the $V^{IV}/V^{V}$ redox couple in NVPF (~4.53 V vs. $Na/Na^+$ at $Na_0$-$Na_1$, **Figure 2d**). Reversible cycling between $Na_1$ and $Na_2$, and between $Na_2$ and $Na_3$ in NVNbPF activates the $Nb^{IV}/Nb^{V}$ and $V^{II}/V^{III}$ redox couples at a voltage of 2.47 and 1.59 V vs. $Na/Na^+$, respectively. The resulting gravimetric energy density of NVNbPF is 428.5 Wh/kg, which is 15.5% lower than that of NVPF.

A commonality of NNbPF and NMoPF is the low value of computed intercalation voltage (0.18 V and 0.28 V vs. $Na/Na^+$, respectively) for the insertion of the fourth Na-ion in the composition range $3 \leq x \leq 4$. To this end, we recomputed the "ground-state" structures of $N_3$MoPF, and $N_4$MoPF, as well as $N_3$NbPF, and $N_4$NbPF, with the same settings implemented by Materials Project.[30] We combined these new data with all the available phases in the respective Na-Mo-P-O-F and Na-Nb-P-O-F quinary composition spaces that we used to derive the thermodynamic stabilities of $N_3$MoPF, $N_4$MoPF, $N_3$NbPF, and $N_4$NbPF. Surprisingly, both $N_3$MoPF and $N_4$MoPF are stable compounds in the Na–Mo–P–O–F phase field, with appreciable formation energies of

~ –2.59 eV/atom, and ~ –2.44 eV/atom, respectively. In contrast, the computed Na–Nb–P–O–F quinary space suggests that both $N_3NbPF$ and $N_4NbPF$ appear highly unstable, with the energy above the convex hull of ~245 meV/atom, and ~318 meV/atom, respectively.

Replacing all V atoms with W, giving NWPF, substantially lowers the voltage plateaus of NVPF. In the Na composition range $0 \leq x \leq 3$, we find three voltage plateaus at ~3.06, ~2.12, and ~1.60 V vs. $Na/Na^+$ for NWPF. This voltage lowering results in a theoretical gravimetric energy density of only 265.9 Wh/kg in NWPF, which is ~47.6% lower than NVPF.

**Sodium-Vacancy and Transition Metal Ordering in $Na_xV_2(PO_4)_2F_3$ Derivatives**

Having established that Mo gives the largest increase in energy density, the remainder of this paper will focus on the structural and kinetic properties of Mo-substituted NVPF systems ($Na_xV_{2-y}Mo_y(PO_4)_2F_3$ with y=0, 1, 2).

**Figure 4a** shows the change in lattice parameters *a* and *c* as a function of the Na content in the NVPF, NMoPF, and NVMoPF systems. **Figure 4b** illustrates the relative volume change of NVPF, NVMoPF, and NMoPF, with the $Na_2$ composition serving as the reference point. All structural parameters are in **Supporting Table S2**.

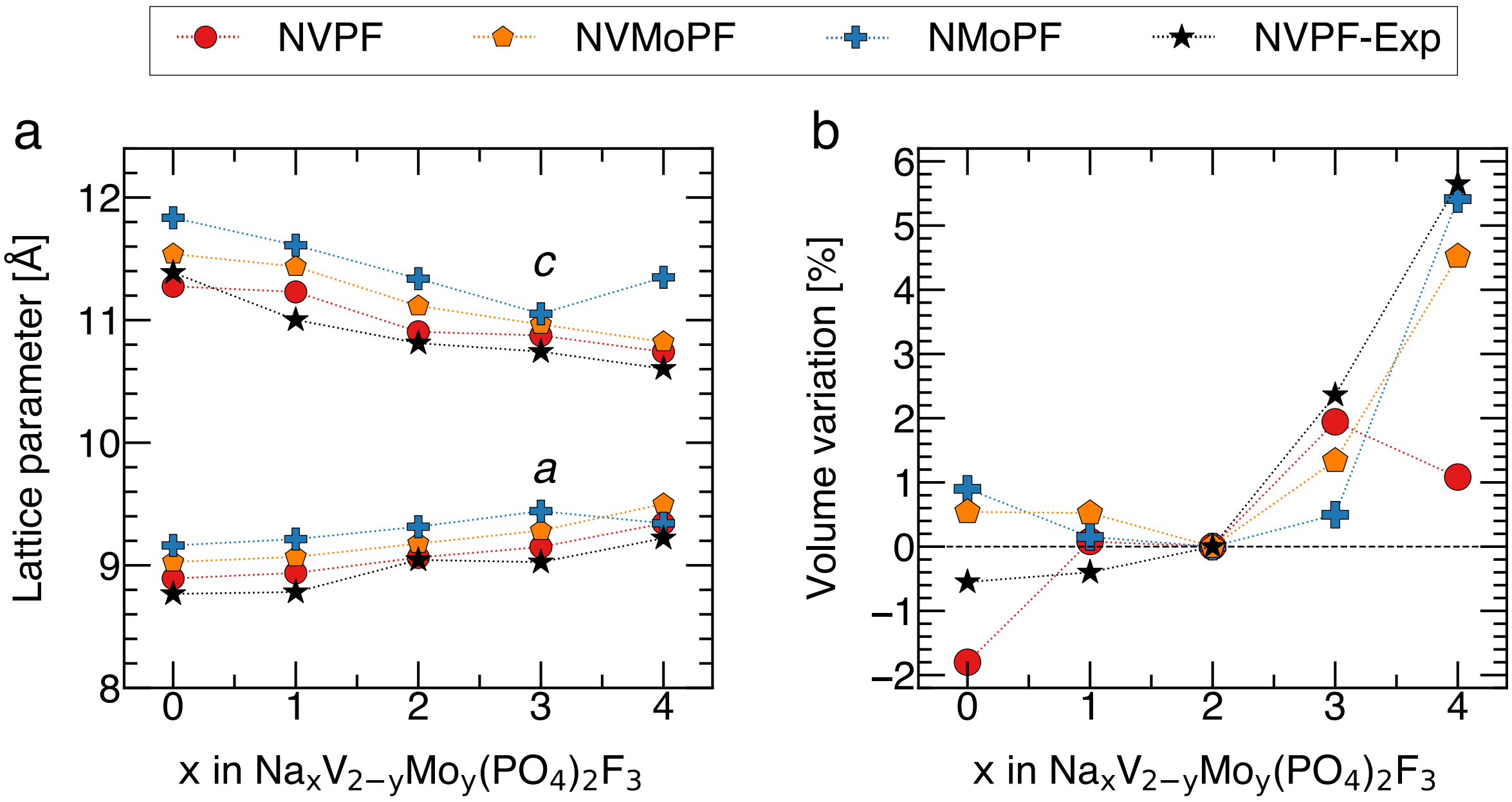

**Figure 4**. Calculated lattice parameters *a* and *c* (panel **a**), and percentual volume variation (panel **b**) for $Na_xV_2(PO_4)_2F_3$ (red circles), $Na_xVMo(PO_4)_2F_3$ (yellow pentagons), and $Na_xMo_2(PO_4)_2F_3$ (blue crosses) systems. The volume variations are benchmarked to the volume of composition $Na_2$, as denoted by the horizontal dotted line in panel **b**. Experimental values (NVPF-Exp)[10,15,26] of lattice parameters and volumes are denoted by black stars in panel **a** and panel **b** Reproduced and adapted with permission under a Creative Commons 4 from Ref. 10. Copyright 2019 Springer Nature. Reproduced and adapted with permission from Ref. 15, 2015 American Chemical Society. Reproduced and adapted with permission under a Creative Commons 4 from Ref. 26. Copyright 2016 Springer Nature). The $Na_2$ composition was chosen as the reference as it is speculated to be the pristine composition in the NVMoPF (V:Mo = 1:1) system, see Figure 2b. Dashed lines are shown for visual guidance.

Quantitatively, the computed values of the lattice parameters agree well with experimental data for NVPF, where the length of *c* generally decreases upon $Na^+$ insertion (upper bundle in **Figure 4a**). This behavior is also consistent for $N_x$VMoPF and $N_x$MoPF, except for $N_4$MoPF, which is accompanied by an increase of the lattice parameter *c*. Such an increase is attributed to the distortion of the bioctahedral units (**Supporting Figure S3**), which is not observed in the fully discharged all-V or mixed V-Mo fluorophosphate materials. The electrochemical cycling of three Na-ions in NVPF (i.e., from $Na_3$ to $Na_0$) leads to a volume variation of ~4%, whereas for NVMoPF and NMoPF, the relative volume change at the same Na composition region is less than 1.5% (**Figure 4b**), indicating good structural stability of these Mo-substituted materials.

Changes of lattice constants and volumes upon (dis)charge of NVPF**,** NVMoPF, and NMoPF are the manifestation of specific spatial Na-vacancy orderings and transition metals arrangements. The Na-vacancy orderings for the ground state structures of NVPF and NVMoPF identified with DFT are displayed in **Figure 5**.

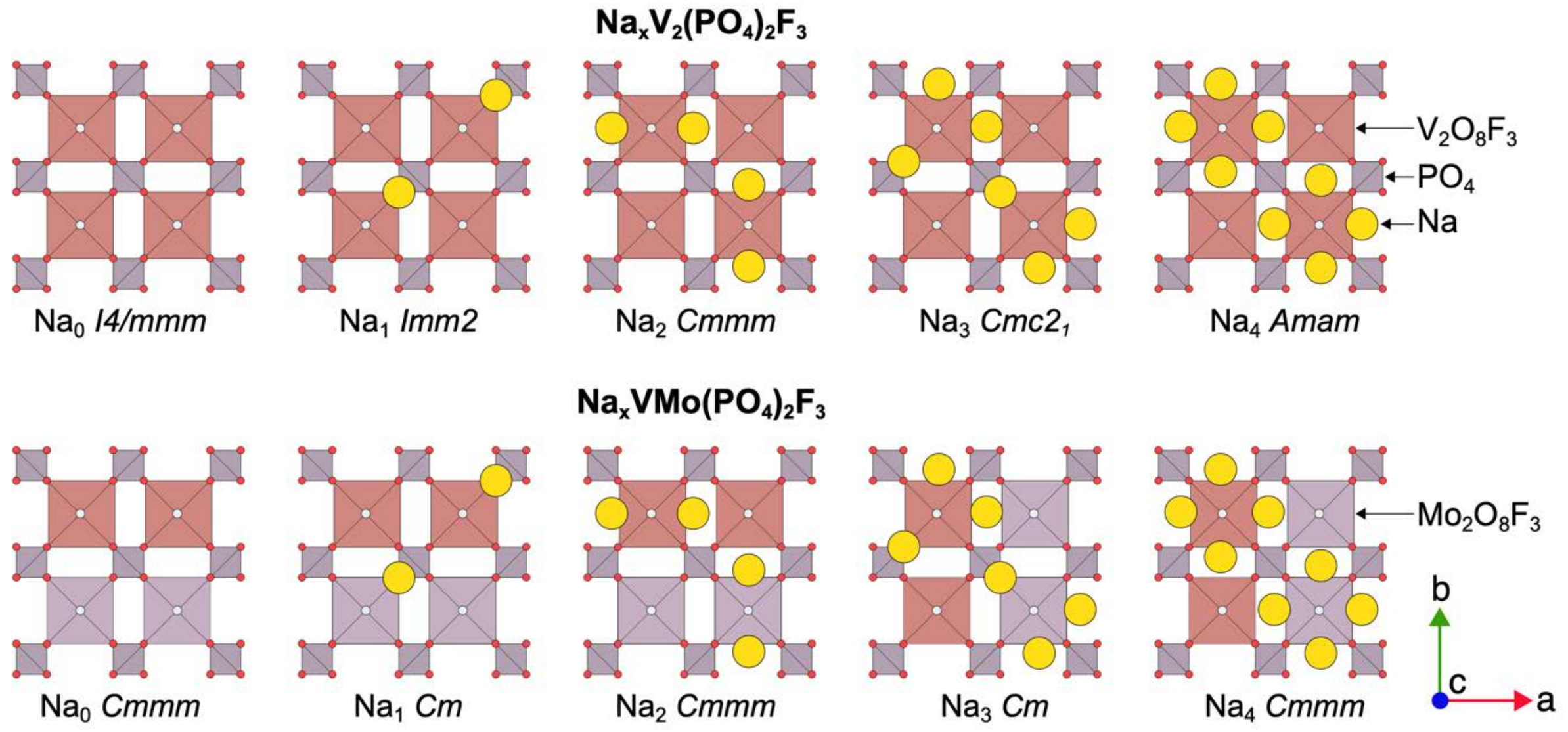

**Figure 5:** Na-vacancy orderings of the ground state structures of $N_xVPF$ (upper part) and $N_xVMoPF$ (lower part) obtained at the GGA+*U* level of theory. Na ions (yellow), V bioctahedral units (red), and Mo bioctahedral units (violet) are shown. Phosphate groups are shown with smaller rectangles in magenta, while oxygen and fluorine atoms are in red and silver, respectively. All structures and space groups correspond to the computed lowest energy structures that form the convex hull (see **Figure 2**).

To develop structure-property relationships in the Mo-containing fluorophosphate, it is important to benchmark our predictions to existing structural studies on the NVPF analog. For the empty vanadium fluorophosphate, Tarascon et al.[10] determined the $N_0VPF$ structure via *ex-situ* synchrotron powder X-ray (PXRD) measurements and found a tetragonal *I*4*/mmm* space group. Our GGA+*U* calculations revealed the same tetragonal *I*4*/mmm* spatial arrangement of atoms for $N_0VPF$.

Bianchini et al.[15] described the $N_1VPF$ structure in the orthorhombic *Cmc2₁* space group, where all Na-ions occupy the Na(1) positions. We predict a ground state $N_1VPF$ structure with the same orthorhombic symmetry (in the *Imm2* space group) with $Na^+$ occupying the Na(3) position. However, we also identified a nearly degenerate structure at $N_1VPF$ with $Cmc2_1$ symmetry, where Na-ions occupy the Na(1) position (see **Supporting Figure S4**) which is only 2.12 meV/atom higher in energy than the ground state structure. The fact that these two Na-vacancy orderings (*Imm2* and $Cmc2_1$) show similar energies may indicate Na exchange between the Na(3) and Na(1) sites, as also observed by previous first-principles calculations.[21]

For $N_2VPF$, Bianchini et al.[15] found a tetragonal structure (*I4/mmm*), which assumes an average distribution of four half-occupied Na positions on the [100] and [010] positions. We found instead an ordered Na-vacancy arrangement, resulting in the orthorhombic symmetry (*Cmmm*). In agreement with Bianchini et al,[15] our ground state at $N_2VPF$ shows that Na-ions occupy either the [100] or the [010] positions. Averaging out the different sodium arrangements of the [100] and the [010] positions from our DFT results can reproduce the same sodium orderings of Bianchini et al.

For $N_3VPF$, Bianchini et al.[15] found the orthorhombic *Amam* space group. Our calculated ground state structure exhibits the orthorhombic $Cmc2_1$ space group. The orthorhombic distortion calculated by our method (b/a=1.003) is in close agreement with the experimental value of 1.002.[24] The remaining discrepancy between the experimental *(Amam)* and calculated ($Cmc2_1$) space groups can be explained by considering that the experimental structure incorporates several different $Na^+$ orderings at room temperature, resulting in partial occupancies of the Na-ions among available positions, whereas our 0 K computational models consider fully ordered structures. Furthermore, we identified five additional structures that are only 4 meV/f.u. or less above the ground state energy (see **Supporting Figure S5**). The experimental structure that can be assumed as an average of these calculated structures coexisting at room temperature is accurately reproduced by our calculations if one considers the entropy scale of ~25 meV/atom at room temperature.

Finally, using the PXRD measurements, Zhang et al. identified the *Amam* space group for the fully discharged compound $N_4VPF$ with all Na(1) and Na(2) positions being fully occupied. Our calculations predict the same orthorhombic symmetry (*Amam*) with the same sodium occupations of Na(1) and Na(2) as observed in the experiment.

In NVMoPF, we found identical Na-vacancy orderings as in the NVPF framework for all sodium compositions (i.e., x = 0, 1, 2, 3, 4). $N_2VMoPF$ arranges in the same orthorhombic symmetry structure as the DFT-predicted ground state of $N_2VPF$, whereas the ground state compositions of NVMoPF at x = 0, 1, 3, 4 are crystallized in different structures compared to NVPF (**Figure 5**). Specifically, the VMoPF, $N_1VMoPF$, $N_3VMoPF$, and $N_4VMoPF$ ground state structures have a lower symmetry than their NVPF analogs, mainly due to the V-Mo arrangements with specific charges.

The different ionic radii of Mo and V and the additional valence electron in Mo will influence the polyanionic framework and the distributions of Na-ions. Hence, we gain more insights into the structural evolution of NVMoPF by analyzing the oxidation states of V and Mo, and the topologies of the bioctahedral units within NVMoPF, as shown in **Figure 6**.

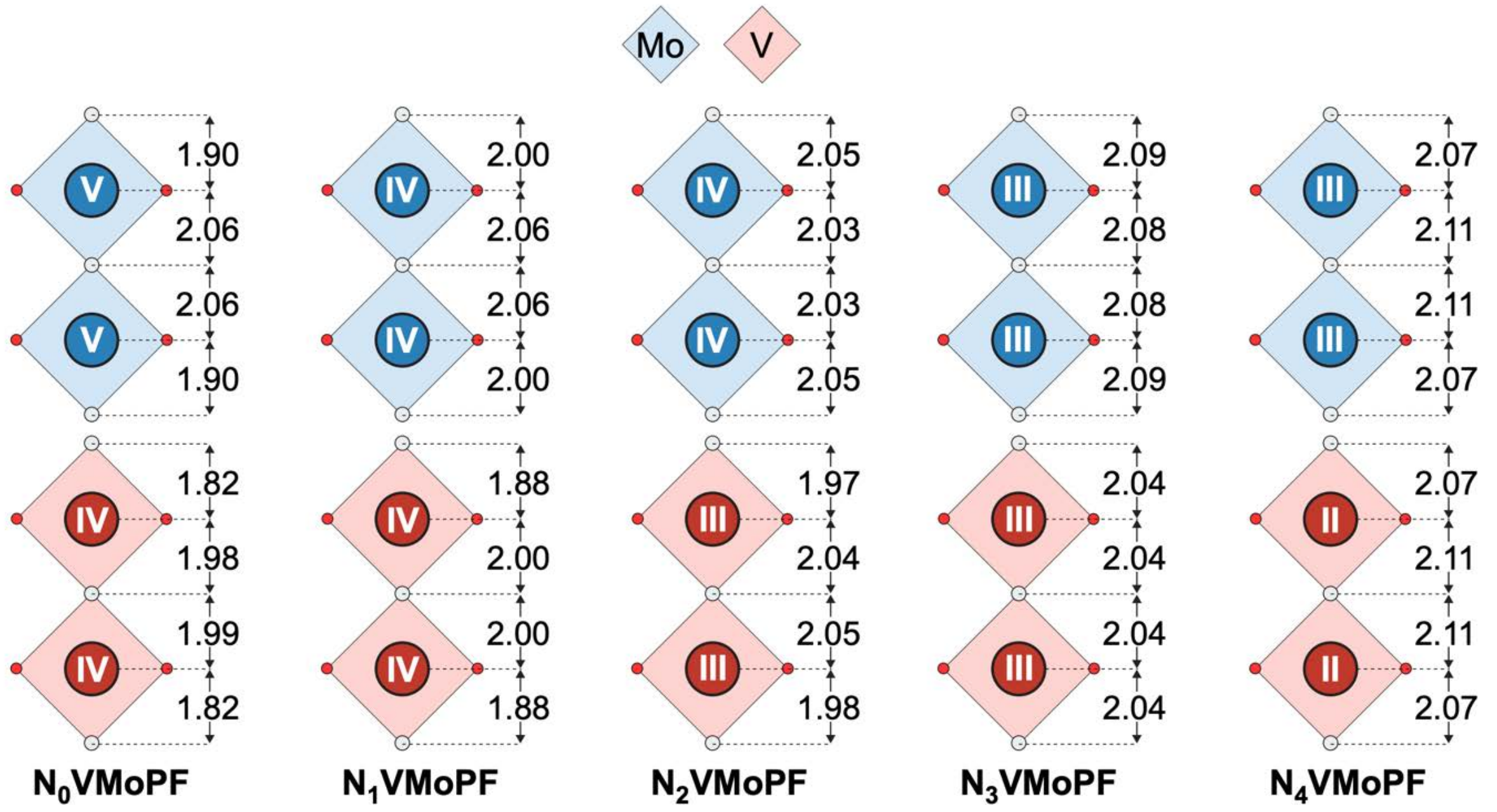

**Figure 6**: Bioctahedral units for $N_xVMoPF$. The oxidation states of V (white text in red circles) and Mo (white text in blue circles) are obtained from DFT+*U* calculations. Bond lengths between the fluorine atoms and the transition metal atoms, either V (F–V) or Mo (F–Mo) are shown in Å. The on-site magnetic moments used to derive the respective oxidation states are ~1.0 $\mu_B$ for $V^{4+}$ and $Mo^{5+}$, ~1.9 $\mu_B$ for $V^{3+}$ and $Mo^{4+}$, and ~2.7 $\mu_B$ for $V^{2+}$ and $Mo^{3+}$.

As it is evident from **Figure 6**, the introduction of Mo into NVPF leads to ordered V-Mo arrangements in NVMoPF, where each bioctahedral unit is composed of either only V ($V_2O_8F_3$) or only Mo ($V_2Mo_8F_3$). Furthermore, within a single bioctahedral unit, both transition metals (i.e., V or Mo) consistently exhibit the same oxidation state, suggesting ordered charges on V/Mo sites. When V and Mo do not exhibit the same oxidation state, i.e., at VMoPF, $N_2VMoPF$, and $N_4VMoPF$, the V ions always show a lower oxidation state than Mo. Specifically, for VMoPF, $N_2VMoPF$, and $N_4VMoPF$, the Mo bioctahedra are in V, IV, and III oxidation states, respectively, whereas the vanadium bioctahedra are in IV, III, and II oxidation states, respectively.

For NVPF, in contrast, our calculations reveal that the two vanadium ions within the same bioctahedral exhibit two different oxidation states for VPF ($V^{IV}$ and $V^{V}$) and $N_4VPF$ ($V^{II}$ and $V^{III}$, see **Supporting Figure S6**), suggesting that NVPF is more prone to charge disordering than NVMoPF. At specific Na compositions, experimental studies have revealed that vanadium may be further disproportionate in $N_1VPF$ (via $2V^{4+} \rightarrow V^{3+}+V^{5+}$). [9,31] Our calculations could not predict this disproportionation reaction in NVPF (see **Supporting Figure S6**), and by extension in NVMoPF.

Clearly, the occurrence of specific V and Mo oxidation states directly influences the structural features of the composition of the electrode materials at different (dis)charge stages. As can be seen in **Figure 6**, the F–V bond lengths are shorter than their corresponding F–Mo bond lengths when the transition metals (TMs) share the same oxidation state, which is consistent with the larger ionic radius of Mo compared to V. We observed a general tendency for the F–TM bonds to elongate with lower oxidation states of the TM, which is evident from an increase in bond length during Na intercalation from VMoPF to $N_4VMoPF$. This bond elongation can be understood from the increase in ionic radius with lower TM oxidation states which leads to longer F–TM bond lengths,[32] and is in line with experimental findings for parent NVPF (see **Supporting Figure S7**).[10,15,24,26] The bond length increase with lower oxidation states is very similar for NVPF and NVMoPF (see **Supporting Figure S7**).

### Sodium-ion Mobility in $Na_xV_2(PO_4)_2F_3$ Derivatives

To understand the extent of the Na-ion mobility in Mo-substituted NVPF derivatives, we performed nudged elastic band (NEB) simulations coupled with GGA*+U* calculations. In these NEBs for NVMoPF, we assumed the V-Mo charge ordering provided by the ground-state $N_2VMoPF$. For parent NVPF, we used the host skeleton of the ground-state $N_3VPF$ to evaluate $Na^+$ mobility. Using these model structures, we studied the effect of Na content on the migration barriers by invoking the high-vacancy limit or charged state (one Na per f.u. of NVMoPF/NVPF) and the low-vacancy limit or discharged state (one vacancy per f.u. of NVMoPF/NVPF). The *ab*-plane view of $Na_2VMo(PO_4)_2F_3$ is shown in **Figure 7a**, with the three crystallographic distinct Na sites (i.e., Na(1), Na(2), and Na(3) as denoted in **Figure 1**) superimposed on the host skeleton.

Two typical $Na^+$ migration pathways with two-dimensional character (i.e., within the *ab*-plane) can be defined in the structure of the NVPF derivatives, namely the intra-unit and inter-unit pathways. The intra-unit pathway adopts a ring-link shape surrounding four nearest-neighbor (NN) Na(1)/Na(2) sites and is indicated by the blue rings in **Figure 7a**. The inter-unit pathway is indicated by the green arrow in **Figure 7a**, and accounts for the Na-ion migration across the intra-unit rings. These two migration pathways can be decoupled into three different $Na^+$ migration paths, denoted as p1, p2, and p3 in **Figure 7a**.[17] Path1 (p1) corresponds to the intra-unit migration which occurs between 2 NN Na(1) and Na(2) sites via an interstitial Na(3) site. Both path2 (p2) and path3 (p3) describe the hops that contribute to the macroscopic inter-unit pathway, where p2 occurs between one Na(1) site and one Na(2) site through two interstitial Na(3) sites, and p3 captures the migration mechanism between two Na(2) sites via the two interstitial Na(3) sites.

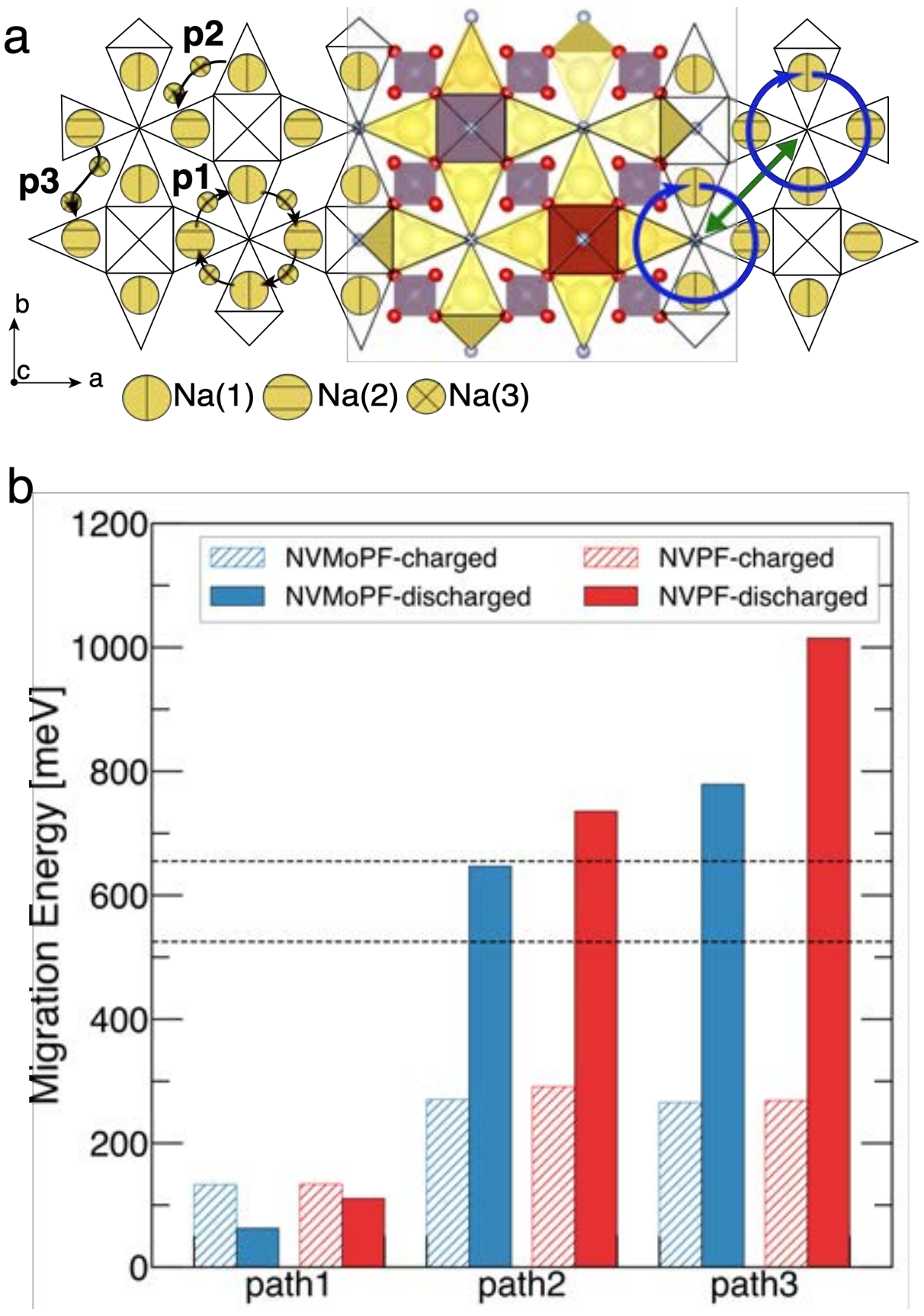

**Figure 7**: Computed $Na^+$ migration barriers in NVPF and NVMoPF. Panel **a** superimposes the three typical Na sites in the NVPF derivatives, i.e., Na(1), Na(2), and Na(3), and schematically represents the three symmetrically distinct $Na^+$ migration pathways, i.e., path1 (p1), path2 (p2), and path3 (p3). p1 corresponds to the intra-unit pathway (blue rings), whereas p2 and p3 contribute to the inter-unit pathway (green arrow). Panel **b** shows the Na-migration barrier energies as calculated by NEBs for the fully charged (x=0) and fully discharged (x=4) NVMoPF (blue) and NVPF (red). The migration energy thresholds of 525 and 650 meV for adequate battery operation in micro-size and nano-size particles, respectively [33] (reproduced and adapted with permission from Ref. 33. Copyright 2015 American Chemical Society), are set by the horizontal lines.

Focusing on the computed migration energies of **Figure 7b**, both NVMoPF, and NVPF show significantly lower migration barriers for the intra-unit pathway p1 than for the inter-unit pathways p2 and p3 at both the discharged and charged limits. Therefore, the inter-unit p2 and p3 pathways appear as the rate-limiting steps for $Na^+$ diffusion in NVMoPF and NVPF. The migration barriers of p2 and p3 increase with the intercalation of Na-ions for both NVPF and NVMoPF. For example, in the discharged limit, NVMoPF shows migration barriers of ~647 and ~779 meV for p2 and p3, respectively, whereas in the charged structure (i.e., high vacancy limit) the migration barriers are as low as ~270 and ~266 meV for p2 and p3, respectively. The higher migration barriers for the discharged limits are attributed to the stronger electrostatic repulsions of nearby Na-ions. Conversely, in the charged limit of NVMoPF the migrating Na-ions benefit from a situation of reduced electrostatic repulsions. Note that the hopping mechanism between two Na(1) sites through two interstitial Na(3) sites can also be approximated by p3 with comparable barrier energies through the same divacancy mechanism (see **Supporting Figure S8**).

The migration barriers for p1 are comparable for NVMoPF and NVPF at both the discharged and charged limits. That is, 63 ~ 133 meV for NVMoPF, and 110 ~ 134 meV for NVPF. However, the migration barriers p2 and p3 were subjected to a further reduction when going from NVPF to NVMoPF. For example, the migration barrier of p3 at the discharged limit is ~1015 meV for NVPF, but only ~779 meV for NVMoPF, corresponding to a ~23% decrease. The Na migration barrier of p2 at the fully discharged limit also decreases by ~12% from 736 meV for NVPF to ~647 meV for NVMoPF. Note that the migration barriers of p2 for NVMoPF in the discharged limit are in the tolerable window of ~525 and ~650 meV, for particle sizes of the active material ranging from micrometer to nanometer scales, respectively.[33] These results suggest optimized kinetic properties and superior $Na^+$ diffusion with sustainable (dis)charge rate [33] of NVMoPF compared to NVPF.

## Discussion

The energy density of the $Na_xV_2(PO_4)_2F_3$ (NVPF) electrode would be increased by ~50% by enabling the extraction of the third Na-ion from $N_1VPF$. Here, we set the goal to identify substitution strategies of transition metals on the vanadium centers in NVPF that may release the third $Na^+$ in sodium fluorophosphate-like electrode materials. Using density functional theory, we inspected the viability of replacing partially or entirely vanadium in NVPF with Mo, Nb, and W.

Relying on our computed voltages (**Figure 2**) at the GGA+*U* level of theory, we have inspected the characteristics of Na intercalation into NVPF substituted compounds. For example, we have investigated the partial (1:1 ratio) or complete substitution of Mo into NVPF. We found that the average voltage from $Na_1$ to $Na_3$ for NVMoPF is ~0.15 V higher than for the theoretically predicted NMoPF. This is mainly attributed to the voltage of the redox couple $V^{III}/V^{IV}$ at $Na_1$-$Na_2$ in NVMoPF, which is ~0.27 V higher than the activated $Mo^{III}/Mo^{IV}$ at $Na_1$-$Na_2$ in NMoPF.

When replacing 50% of V with Mo, the voltage plateau associated with the $Na^+$ extraction from x = 1 to x = 0 is lowered from 4.53 V vs. $Na/Na^+$ in NVPF to 3.93 V vs. $Na/Na^+$ in NVMoPF by activating the $Mo^{IV}/Mo^{V}$ (instead of the $V^{IV}/V^{V}$) redox couple. Therefore, we speculate that the extraction of the third $Na^+$ is thermodynamically possible for the NVMoPF positive electrode using the most common electrolytes.[27] The extraction of the third $Na^+$ in NVMoPF results in an improved gravimetric capacity of 173.4 mAh/g and a predicted energy density of 577.3 Wh/kg for NVMoPF (see **Supporting Table S1**). This energy density represents a substantial increase (~13.9%) in comparison to NVPF (i.e., 507 Wh/kg as observed in experiments.)[15] Since GGA+*U* underestimates the voltage of NVPF (vide supra) at each step, the actual increase in energy density for NVMoPF is expected to be more substantial. Specifically, using the theoretically computed (rather than the experimentally reported) energy density of 439 Wh/kg for NVPF as the reference, we predict an increase in energy density of ~31.5% for NVMoPF. An in-depth analysis of the Mo-V spatial arrangements and oxidation states in NVMoPF suggests that Mo and V centers in the mixed fluorophosphate electrodes act entirely independently. Indeed, a closer look at the Mo-V arrangement in the stable ground-state ordering (**Figure 6**) may suggest a tendency for charge ordering of Mo and V bioctahedral units.

We contemplated the possibility of replacing all the V content in NVPF with Mo. Notably, the voltage associated with the full desodiation from $N_1$MoPF at 3.94 V vs. Na/$Na^+$ is similar to $N_1$VMoPF (~3.93 V vs. Na/$Na^+$). An increased capacity resulting from extracting the third $Na^+$ from $N_1$MoPF gives an energy density of ~511.1 Wh/kg for NMoPF (see **Supporting Table S1**) —accounting for an increase of ~0.8% for the experimental energy density of NVPF (~16.4% relative to the computed energy density of NVPF). Our results suggest that this electrochemical process ($N_1$MoPF ~ $N_0$MoPF) may be accessible in NMoPF electrode material.

Furthermore, we investigated the addition of Nb in NVPF (in a 1:1 ratio), as well as the full replacement of V with Nb. Unexpectedly, in NVNbPF the extraction of the third Na-ion from $N_1$VNbPF to the empty VNbPF activates the $V^{III}/V^{IV}$ redox couple at ~3.29 V vs. Na/$Na^+$, which is more than 1 V lower than in the parent NVPF at x = 0 ~ 1 via $V^{IV}/V^{V}$ (~4.53 V vs. Na/$Na^+$). Considering the reversible extraction of three Na-ions, we estimated that NVNbPF has a theoretical gravimetric energy density of ~428.5 Wh/kg, and 15.5% lower than NVPF. Furthermore, the estimated theoretical energy density of NNbPF (x = 0 ~ 3) is ~301.2 Wh/kg, resulting in a decrease of 40.6% compared with NVPF. Therefore, both NVNbPF and NNbPF do not chart as promising candidates for Na-ion battery cathodes.

We also investigated the effect of adding W into NVPF. Since W is a much heavier element (183.8 g/mol) than V, Nb, and Mo (50.9, 92.9, and 96.0 g/mol, respectively), the computed theoretical capacity of NWPF is only ~39.2 mAh/g per Na, which is substantially lower than the ~64.2, ~52.8, and ~53.4 mAh/g (per Na) for NVPF, NMoPF, and NNbPF electrode materials, respectively. Therefore, the predicted energy density using our voltages (**Figure 3**) for NWPF is only 265.9 Wh/kg, corresponding to a 47.6% decrease in comparison with NVPF (**Table S1** of SI). **Figure 8** summarizes the predicted voltages and energy densities for each intercalation step in $Na_xV_{2-y}M_y(PO_4)_2F_3$ at various substitution levels of Mo, Nb, and W.

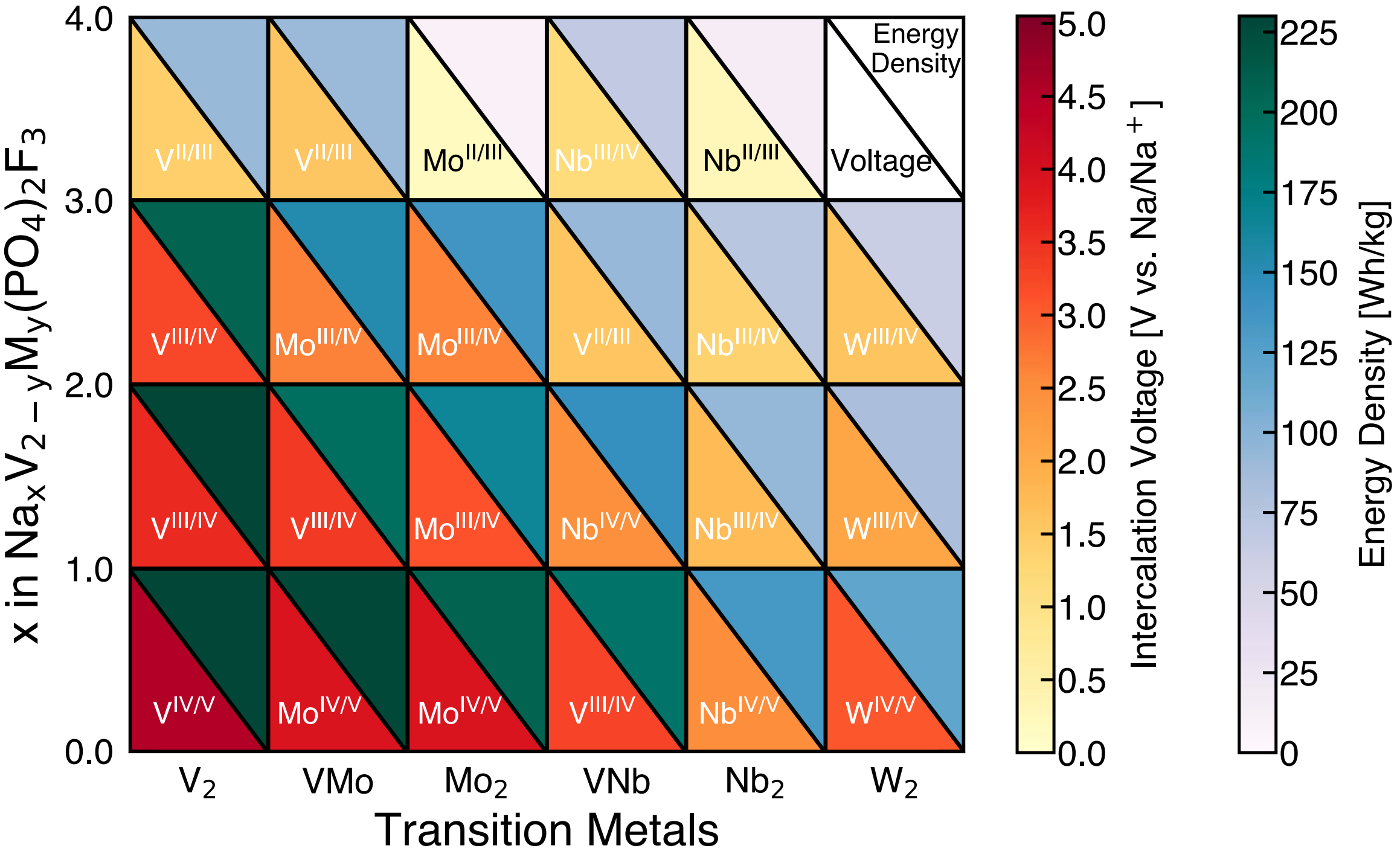

**Figure 8.** Map of computed intercalation voltages and energy densities for $Na_xV_{2-y}M_y(PO_4)_2F_3$ systems. The x-axis represents the different transition metal combinations of V and M, where M = Mo, Nb, and W. The y-axis represents the Na concentration x from x = 0 to x = 4. Each square is divided into lower and higher triangles, which are mapped onto the specific values of intercalation voltage and energy density, respectively. The redox couple for each intercalation step is superimposed on the lower triangle of each square.

To elucidate the feasibility of extracting the third Na-ion from $N_1$VMoPF, we have studied the kinetic properties of NVMoPF. We established that the rate-determining steps are the inter-unit pathways, i.e., the p2 and p3 pathways (**Figure 7**) that contribute to the macroscopic diffusion of Na ions in fluorophosphate-like electrode materials. We found that the $Na^+$ migration energy barriers at the charged limit are ~270 meV (p2) and ~266 (p3) meV for NVMoPF, and ~291 meV (p2) and ~269 meV (p3) for NVPF, which are all below the tolerance limits of 525 and 650 meV (see **Figure 7**). These low values for migration barriers at low Na composition regions suggest no obvious kinetic limitations imposed on the migrating Na-ion in NVPF and NVMoPF.

Nevertheless, it has been shown experimentally that NVPF exhibits a rapid decline in $Na^+$ diffusivity as x approaches 1,[20] which was attributed to the high energies required to create vacancies in NVPF at low Na concentration. Therefore, we attribute the difficulty in extracting the third Na-ion (i.e., from x = 1 to x = 0) to the strong Na-vacancy orderings at x = 1. Such a strong interplay of Na-vacancy at low Na concentration hinders the creation of vacancies that are needed for Na-ion mobility, resulting in a higher migration energy penalty to promote $Na^+$ diffusion. Consequently, the $Na^+$ migration is impractical to occur at x = 1. As a measure for the Na-vacancy ordering at low Na content (charged state), one may examine the voltage step between x = 0~1 and x = 1~2. For NVPF (**Figure 2d**), we observe a steep decrease of ~0.93 V in the intercalation voltage (i.e., from 4.53 to 3.60 V vs. $Na/Na^+$), suggesting a rather strong gain in stabilization induced by the effect of Na-vacancy ordering at $N_1VPF$.

Interestingly, the same voltage step in $N_1VMoPF$ is only 0.53 V (**Figure 2e**), corresponding to a decrease of 43% compared to $N_1VPF$. A milder decrease in intercalation voltage is driven by mixing V with Mo, which weakens the strong tendency of Na-vacancy ordering, thereby facilitating Na-ion migration.[21,34]

Importantly, Dacek et al. showed that vacancy formation energy in $N_1VPF$[21] is only favorable if sodium extraction occurs at the same voltage as the equilibrium oxidation potential. In practice, the high intercalation voltage of $N_0VPF$–$N_1VPF$, well above the electrolyte stability window (~4.5 V vs. $Na/Na^+$) leads to a practical extraction potential that is lower than the equilibrium potential, [35] resulting in poor kinetics for the NVPF system with high defect formation energies. According to our predictions, the extraction of the third Na-ion in NVMoPF occurs at a voltage that is below the stability window of most electrolytes. This lower voltage may allow the third Na-ion to be extracted at the equilibrium oxidation voltage, and therefore favorable kinetics are expected for the NVMoPF electrode.

Overall, both the lower migration barrier and weaker Na-vacancy orderings in NVMoPF as compared to NVPF indicate an improvement of the accessible (de)intercalation capacity in the charged states. Specifically, the theoretical capacity of NVMoPF is evaluated to be ~174 mAh/g, corresponding to insertion/extraction between $N_0VMoPF$ and $N_3VMoPF$, which is 35% higher than 128 mAh/g of NVPF (at x = 1 ~ 3, see **Supporting Table S1**).

## Conclusions

We investigated the substitution of vanadium (V) by molybdenum (Mo), niobium (Nb), and tungsten (W) in $Na_xV_2(PO_4)_2F_3$ (NVPF) to improve the gravimetric energy density of this promising electrode material for Na-ion batteries. Using first-principles calculations based on density functional theory, we elucidated the electrochemical and structural properties of NVPF derivatives at different substitution levels.

Partial substitution of 50% V with Mo reduces the voltage plateau for extracting the third $Na^+$ from ~4.5 to ~3.9 V vs. $Na/Na^+$, thus enabling the practical Na extraction from $Na_1VMo(PO_4)_2F_3$ and increasing the capacity from ~128 to ~174 mAh/g. Structural analysis revealed that $Na_xVMo(PO_4)_2F_3$ (NVMoPF) undergoes a minimal volume change of only 1.3% upon reversible Na extraction, suggesting good structural stability during cycling. An in-depth analysis of the Na-ion migration barriers showed that NVMoPF may have superior kinetic properties over NVPF, which was attributed to the disrupted $Na^+$ orderings enabled by the extra entropy of mixing Mo with V in the fluorophosphate material. The mixing of Mo with V may deliver a gravimetric energy density of ~577.3 Wh/kg for $Na_xVMo(PO_4)_2F_3$ versus 507 Wh/kg for $Na_xV_2(PO_4)_2F_3$, corresponding to an increase of 13.9%. The high practical energy density combined with superior kinetics make NVMoPF a promising positive electrode material for Na-ion batteries.

## Methods

We used the density functional theory (DFT)-based Vienna *Ab initio* Simulation Package (VASP) [36,37] to evaluate the structural stability, total energy, thermodynamic and electronic properties of $Na_xV_{2-y}M_y(PO_4)_2F_3$ positive electrode materials ($0 \leq x \leq 4$, y = 2, 1.5, 1, 0.5, and 0, and M = V, Mo, Nb, and W). The spin-polarized generalized gradient approximation (GGA)-type Perdew–Burke–Ernzerhof (PBE) functional was used to approximate the unknown exchange-correlation contribution in DFT.[38] We also added a Hubbard *U* correction to address the strong on-site Coulomb correlation effects of the *3d* electrons of V, the *4d* electrons of Nb and Mo, and the *5d* electrons of W.[39,40] The effective *U* parameters applied in this study are 3.1 eV for V, 2.0 eV for Nb, 3.5 eV for Mo, and 4.0 eV for W.[41] A value of *U* of 3.1 eV for V was chosen because it gives reliable formation energies and redox reactions of V-based compounds.[17,21,42,43]

The projected augmented wave (PAW) potentials were used for the description of core electrons,[44] namely, Na 08Apr2002 $3s^13p^0$, P 17Jan2003 $3s^23p^3$, O 08Apr2002 $2s^22p^4$, F 08Apr2002 $2s^22p^5$, V_pv 07Sep2000 $3p^63d^44s^1$, Mo_pv 08Apr2002 4p5s4d, Nb_pv 08Apr2002 4p5s4d, and W_pv 06Sep2000 5p6s5d. Valence electrons were represented using plane waves up to an energy cutoff of 520 eV. In addition, a $\Gamma$-centered Monkhorst-Pack[45] *k*-point mesh with 25 subdivisions along each reciprocal lattice vector was applied to all structures. The total energy of each structure was converged to within $10^{-5}$ eV/cell, and atomic forces within $10^{-2}$ eV/Å.

To explore the Na (de)intercalation behaviors for all the selected $Na_xV_{2-y}M_y(PO_4)_2F_3$ systems, we used the experimentally-observed $Na_3V_2(PO_4)_2F_3$/$Na_4V_2(PO_4)_2F_3$ phases [26] as the starting structures to study Na removal with different degrees of substitution of transition metals (TMs, i.e., M = Mo, Nb, and W in $Na_xV_{2-y}M_y(PO_4)_2F_3$). Specifically, in NVPF we substituted 25%, 50%, 75%, and 100% of V by Mo, or 50% and 100% of V by Nb, or 100% of V by W, resulting in the chemical formulas $Na_xV_{2-y}Mo_y(PO_4)_2F_3$, $Na_xV_{2-y}Nb_y(PO_4)_2F_3$, and $Na_xW_2(PO_4)_2F_3$ with y=0, 0.5, 1.0, 1.5, 2.0 for Mo, and y=1.0, 2.0 for Nb, respectively. For each of the selected systems, we varied the Na content x in the range of $0 \leq x \leq 4$ and with a step $\Delta x$ = 0.5 or 1.0. The possible Na-vacancies and transition metal configurations were enumerated using the pymatgen package.[46] At each Na content x of the specific $Na_xV_{2-y}M_2(PO_4)_2F_3$ system, we chose a maximum of 100 candidate structures with the lowest Ewald energy,[47] based on integer charges (i.e., Na = +1, P= +5, O = –2, F = –

1) and variable charges on the TM (+2 ~ +5) to limit the number of possible candidate structures and keep our calculations computationally tractable. DFT calculations were performed on the primitive cell (2 formula units of $Na_xV_{2-y}M_y(PO_4)_2F_3$ with 30+2x atoms) and supercells (2×1×1) of these enumerated configurations. The oxidation states of the TM-ions were derived by tracking the magnetic moments on the TMs upon structural and electronic relaxation.

The (de)intercalation process for the $Na_xV_{2-y}M_y(PO_4)_2F_3$ positive electrode material can occur through the reversible $Na^+$ insertion/extraction according to the redox Equation (1):

$$\mathrm{Na_xV_{2-y}M_y(PO_4)_2F_3 + (z-x)Na^+ + (z-x)e^-} \overset{-\Delta G}{\longleftrightarrow} \mathrm{Na_zV_{2-y}M_y(PO_4)_2F_3} \qquad (1)$$

where x and z represent the initial and final Na content, and $\Delta G$ is the change of Gibbs free energy at 0 K, which can be approximated by the DFT energy. The formation energies ($\mathrm{E_f(x)}$), which are corresponding to the phase stability of $Na_xV_{2-y}M_y(PO_4)_2F_3$ with respect to the DFT-energies of fully charged ($Na_0V_{2-y}M_y(PO_4)_2F_3$) and discharged phase ($Na_4V_{2-y}M_y(PO_4)_2F_3$), can be computed using Equation (2):

$$\mathrm{E_f(x) = E\big[Na_xV_{2-y}M_y(PO_4)_2F_3\big] - \left(\frac{4-x}{4}\right) E\big[Na_0V_{2-y}M_y(PO_4)_2F_3\big] - \left(\frac{x}{4}\right) E\big[Na_4V_{2-y}M_y(PO_4)_2F_3\big]} \qquad (2)$$

Note that we approximate the 0 K mixing energies using formation energies, neglecting the *pV* and entropic contributions. The intercalation voltage for the redox reaction of Equation (1) can be approximated using Equation (3):

$$V = -\frac{\Delta G}{(\mathrm{z}-\mathrm{x})F}$$
$$= -\frac{E\left[\mathrm{Na_zV_{2-y}M_y(PO_4)_2F_3}\right] - \left(E\left[\mathrm{Na_xV_{2-y}M_y(PO_4)_2F_3}\right] + (\mathrm{z}-\mathrm{x})\mu_{Na}\right)}{(\mathrm{z}-\mathrm{x})F} \quad (3)$$

where $\mu_{Na}$ is the chemical potential of bulk Na metal and $F$ is the Faraday constant. The corresponding theoretical capacity ($C$) for this intercalation step can be evaluated using Equation (4):

$$C = \frac{(\mathrm{z}-\mathrm{x})F}{m} \quad (4)$$

where *m* is set as the mass of the "pristine" $Na_3V_{2-y}M_y(PO_4)_2F_3$ composition to normalize the gravimetric specific capacity (mAh/g).

To investigate the Na ion migration in $Na_xVMo(PO_4)_2F_3$ (NVMoPF), and $Na_xV_2(PO_4)_2F_3$ (NVPF) systems, we employed the nudged elastic band (NEB) method,[48–50] with the same DFT settings discussed above. We reduced the *k*-point sampling to the $\Gamma$-point. For both NVMoPF and NVPF, we did NEB at the fully charged/desodiated (x = 0) and fully discharged/sodiated (x = 4) limits with the fixed transition metal (i.e., V, Mo) skeleton at these 2 limits, which are corresponding to the host skeleton of the V-Mo and V-V arrangements at the ground-state of $Na_2VMo(PO_4)_2F_3$ and $Na_3V_2(PO_4)_2F_3$, respectively. In this way, we could approximate the topotactic $Na^+$ migration during Na (de)intercalation from/into the ground-state structures.

## Associated content

The Supporting Information contains the following items: the formation energies and intercalation voltages for $NV_{2-y}Mo_yPF$ with y=0.5 and 1.5; formation energies for NVNbPF, NNbPF, and NWPF; bioctahedra units with angular data for $N_4VPF$, $N_4VMoPF$, and $N_4MoPF$; Na-vacancy orderings for near-degenerate structures of $N_1VPF$ and $N_3VPF$; bioctahedral units with oxidation states and bond lengths for NVPF and NMoPF; TM–F bond length changes for NVPF, NVMoPF, and NMoPF; NEB energy paths for NVPF and NVMoPF; summary of all intercalation properties in terms of voltages, capacities, and energy densities for NVPF, NVMoPF, NMoPF, NVNbPF, NNbPF, $NV_{1.5}Mo_{0.5}PF$ and $NV_{0.5}Mo_{1.5}PF$; tabulated space group and lattice parameters for $NV_{2-y}Mo_yPF$ with y=0, 0.5, 1.0, 1.5, and 2.0.

## Notes

The authors declare no competing financial interest.

## Acknowledgments

P.C. acknowledges funding from ANR-NRF NRF2019-NRF-ANR073 Na-MASTER. P.C. acknowledges funding from the National Research Foundation under his NRF Fellowship NRFF12-2020-0012.

**Graphic for Table of Content**

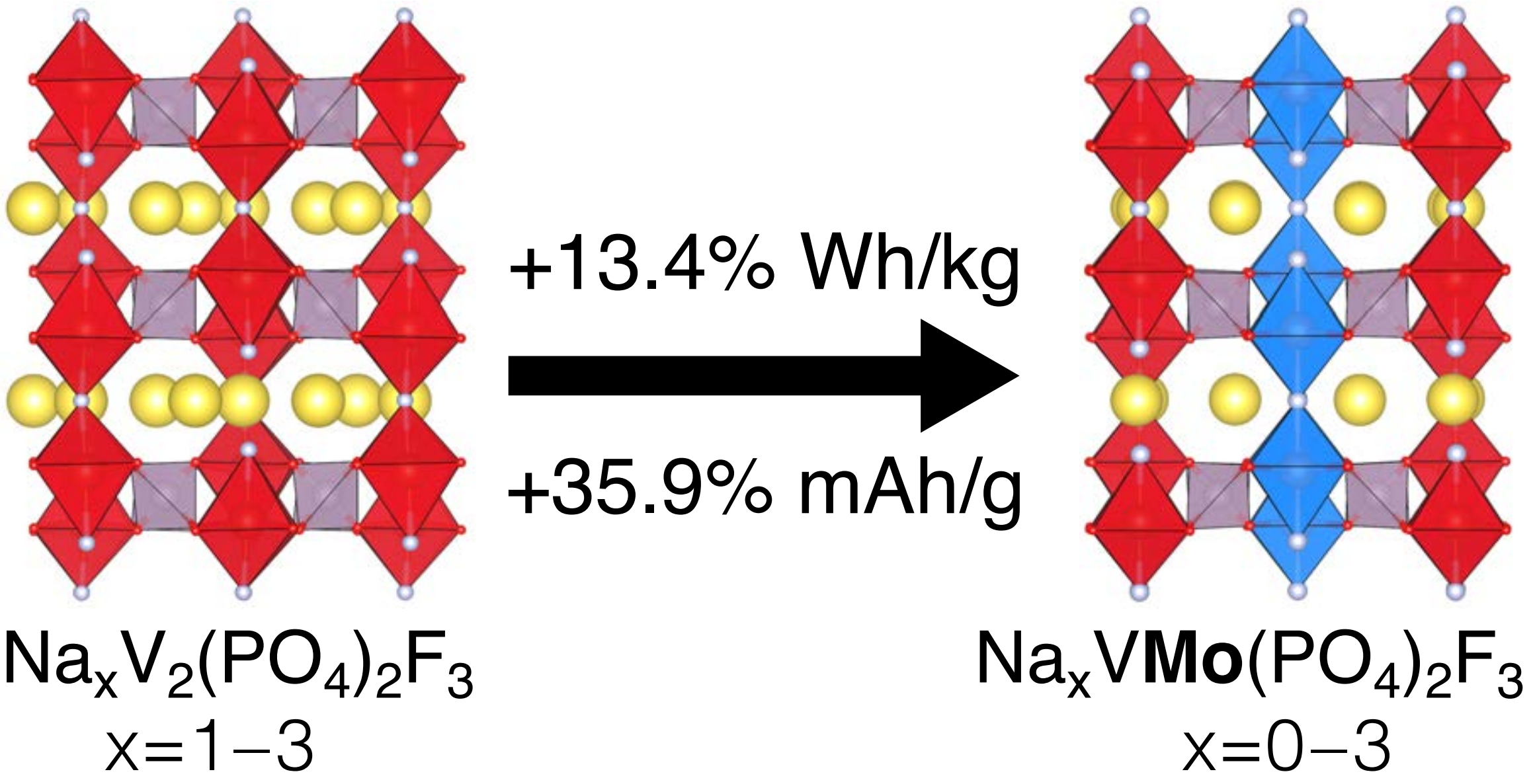

## —Supporting Information—

# Unlocking the Inaccessible Energy Density of Sodium Vanadium Fluorophosphate Electrode Materials by Transition Metal Mixing

Stephanie C. C. van der Lubbe[1#], Ziliang Wang[1#], Damien K. J. Lee[2], and Pieremanuele Canepa[1,2*]

[1]Department of Materials Science and Engineering, National University of Singapore, 9 Engineering Drive 1, 117575, Singapore

[2]Department of Chemical and Biomolecular Engineering, National University of Singapore, 4 Engineering Drive 4, 117585, Singapore

[#]These authors contributed equally to this work

*pcanepa@nus.edu.sg

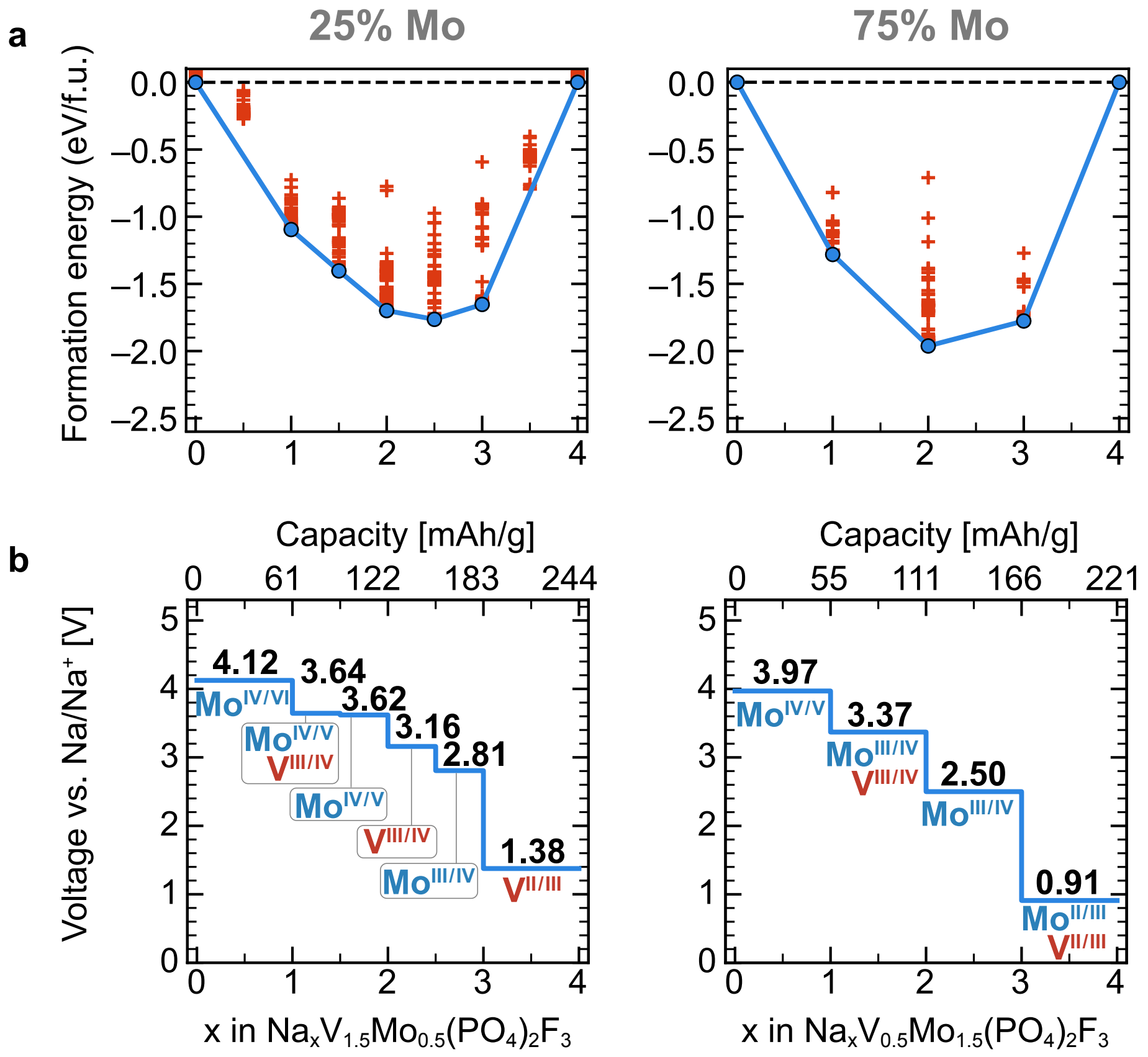

**Supporting Figure S1:** Panel **a** shows the formation energies as computed with DFT as a function of the Na concentration x for $Na_xV_{1.5}Mo_{0.5}(PO_4)_2F_3$ (left) and $Na_xV_{0.5}Mo_{1.5}(PO_4)_2F_3$ (right). The unstable orderings are indicated by the red + symbol, whereas the stable orderings forming the convex hull are displayed as a blue dot with black edge. Panel **b** shows the computed intercalation voltages vs. $Na/Na^+$ as a function of the Na concentration x for $Na_xV_{1.5}Mo_{0.5}(PO_4)_2F_3$ (left) and $Na_xV_{0.5}Mo_{1.5}(PO_4)_2F_3$ (right). All capacities are calculated by using the molar mass at x=3. The active redox couples are indicated in red for V and in blue for Mo. Data obtained at GGA+$U$ level of theory. The computed energy densities are given in **Supporting Table 1**.

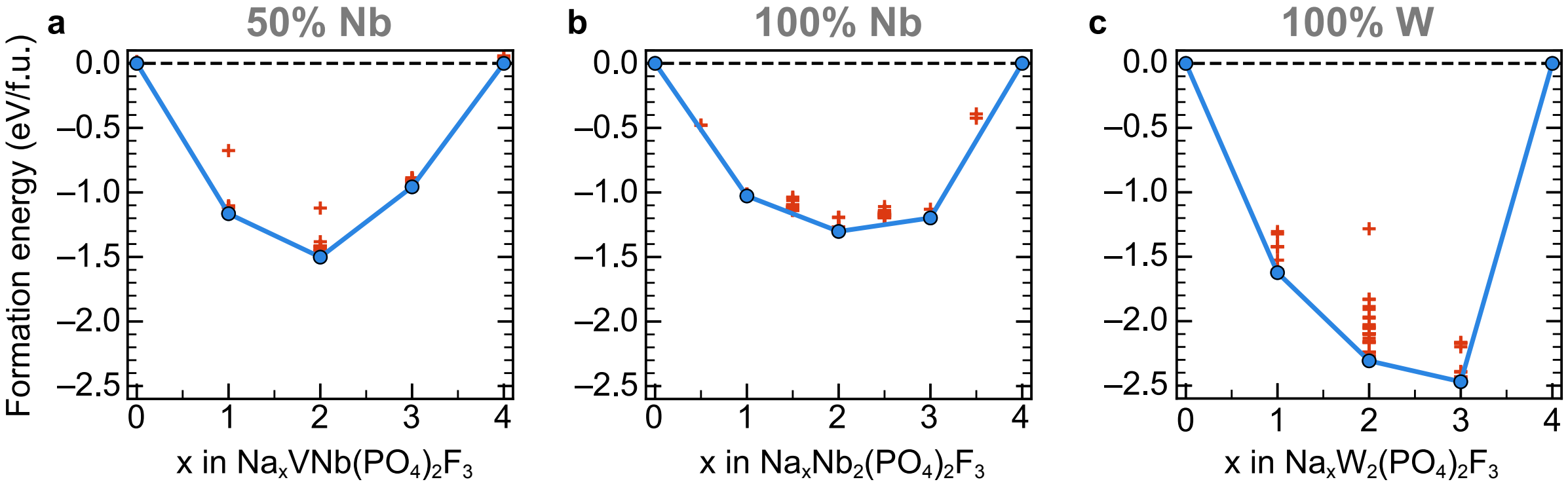

**Supporting Figure S2:** Panels **a**, **b** and **c** show the formation energies as computed with GGA+*U* as a function of the Na concentration x for $Na_xVNb(PO_4)_2F_3$, $Na_xNb_2(PO_4)_2F_3$, and $Na_xW_2(PO_4)_2F_3$, respectively. The unstable orderings are indicated by red crosses, whereas the stable orderings forming the convex hull are displayed as blue dots.

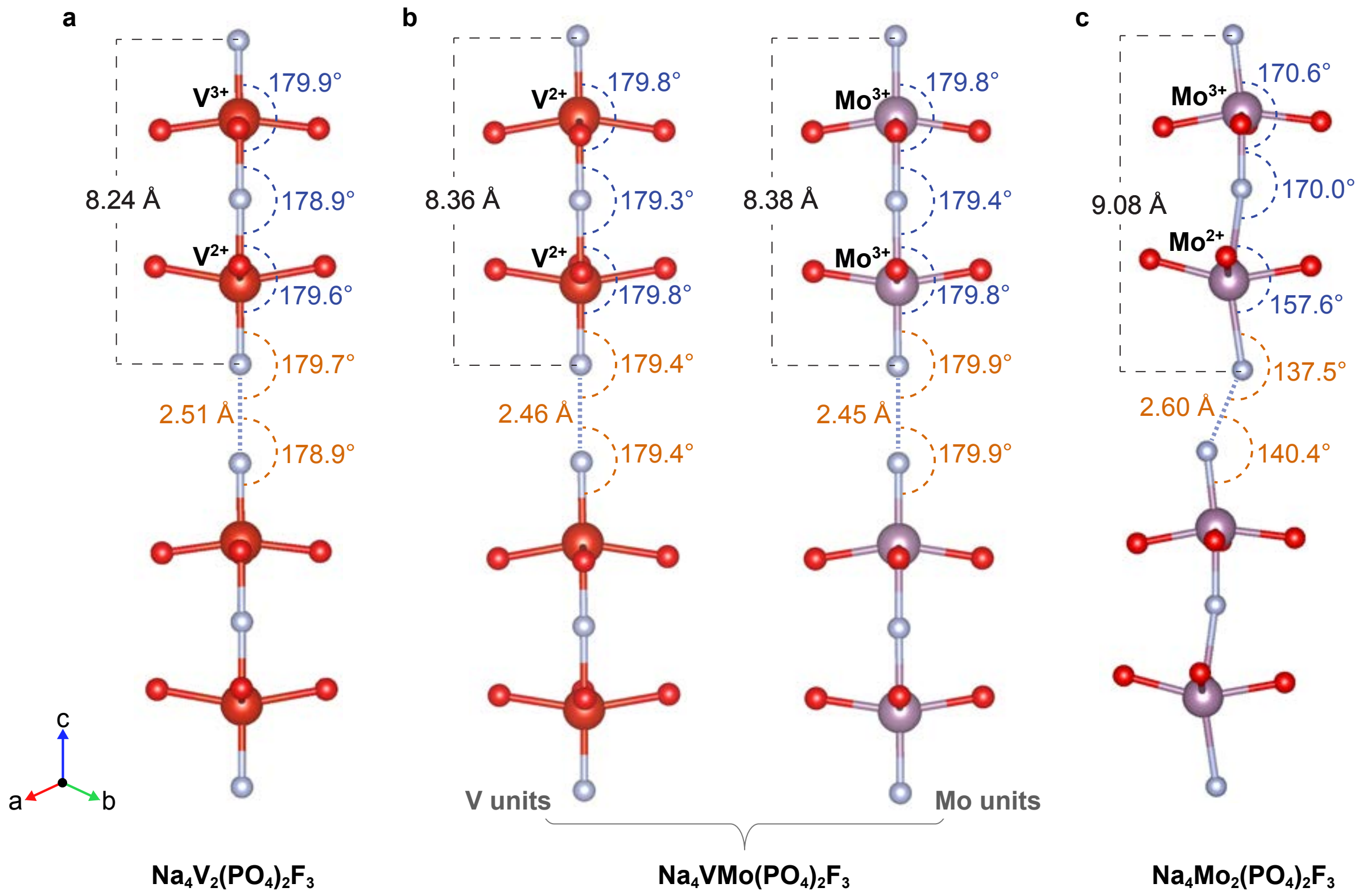

**Supporting Figure S3:** Structure of the **a:** $Na_4V_2(PO_4)_2F_3$, **b:** $Na_4VMo(PO_4)_2F_3$, and **c**: $Na_4Mo_2(PO_4)_2F_3$ bioctahedra with relevant bond lengths and angles given in Angstrom and degrees. Sodium atoms are omitted for visual clarity. The NVPF and NVMoPF bioctahedra have similar structural features with comparable bond lengths and a near-linear framework. In contrast, the NMoPF structure has longer bond lengths and shows a strong deviation from linearity. Data obtained at GGA+*U* level of theory.

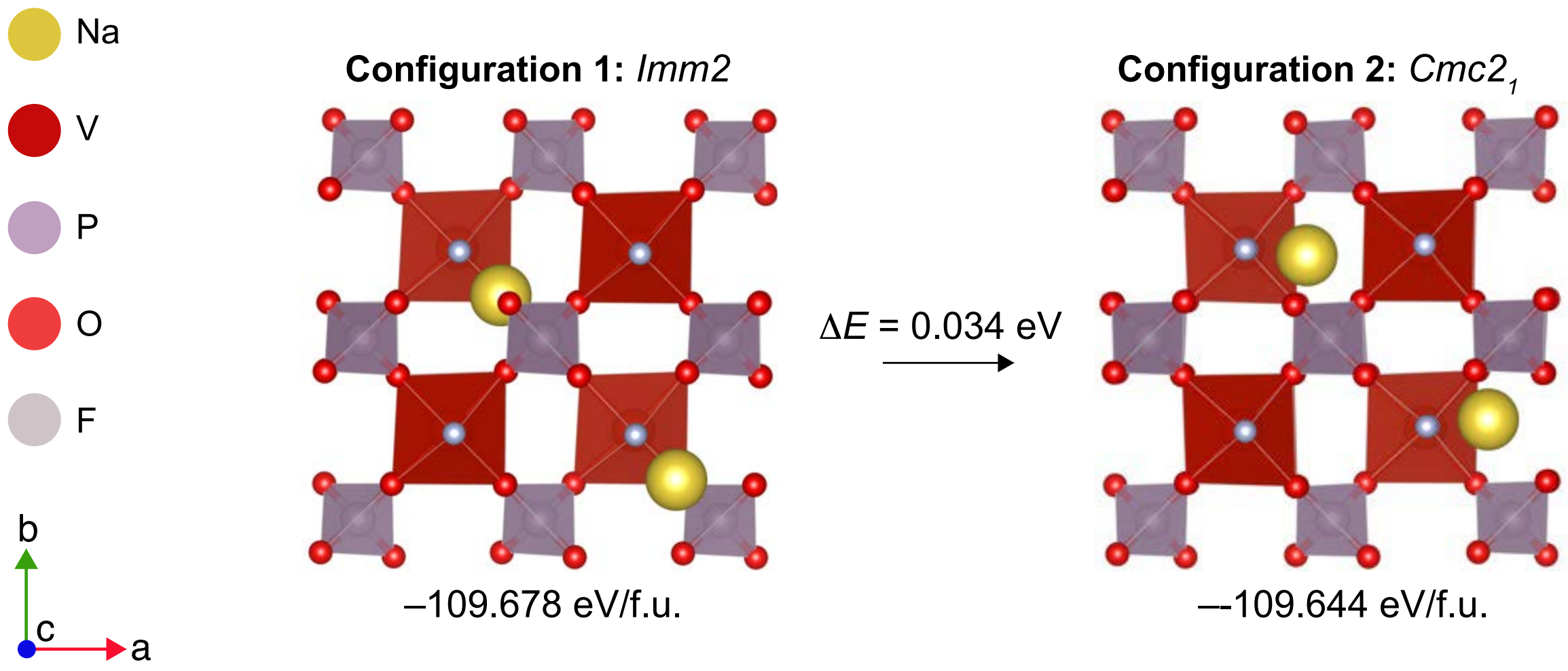

**Supporting Figure S4:** Na-vacancy orderings for $NaV_2(PO_4)_2F_3$. **Configuration 1:** ground state structure with *Imm2* symmetry. **Configuration 2:** structure with $Cmc2_1$ symmetry, which is 0.034 eV/f.u. higher in energy than configuration 1. Data obtained at GGA+*U* level of theory.

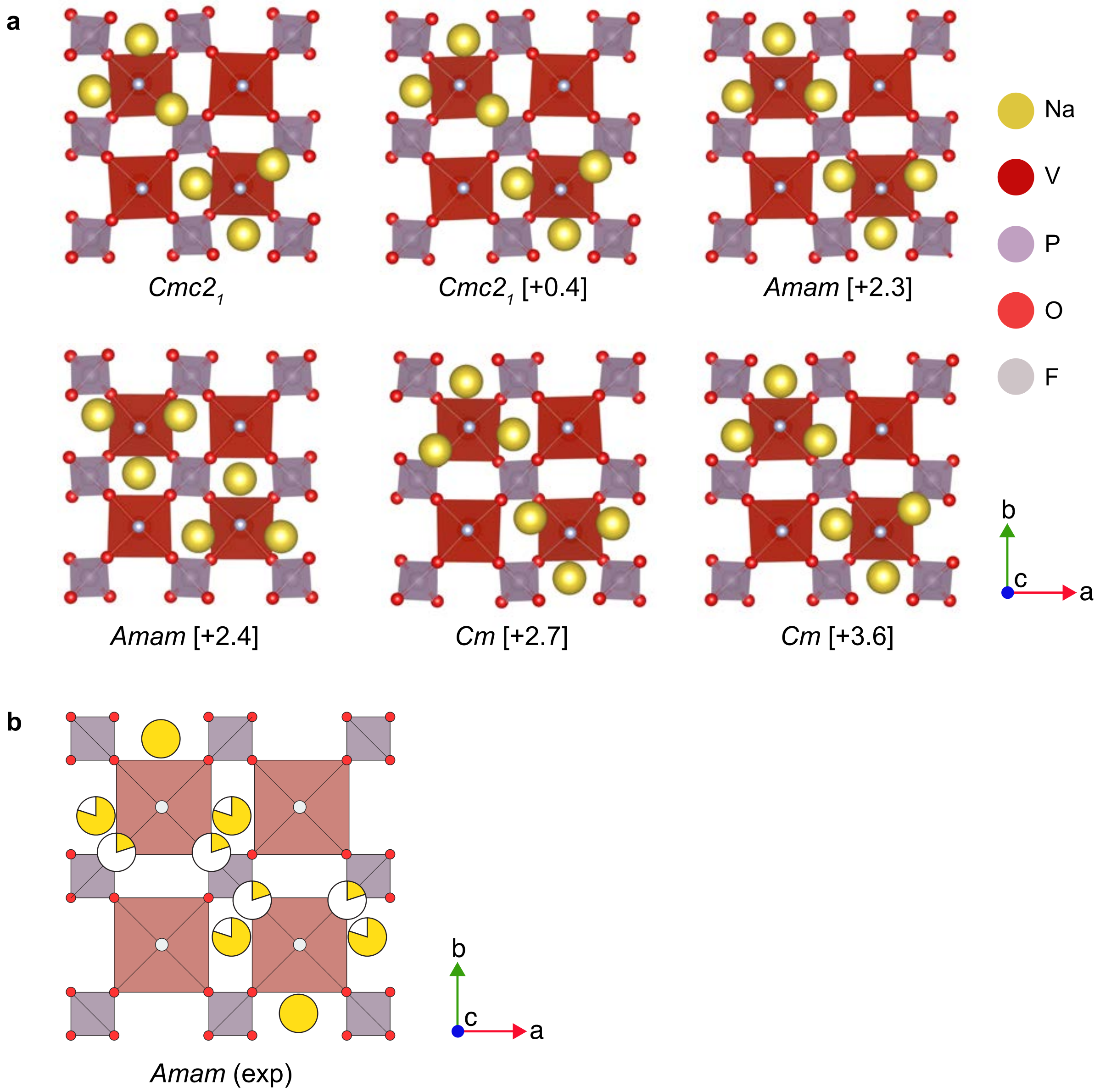

**Supporting Figure S5:** Na-vacancy orderings for $Na_3V_2(PO_4)_2F_3$. Panel **a** shows the six lowest energy structures as obtained in this work using GGA+*U*, with energy differences from the ground state structure shown in square brackets in meV/f.u. Panel **b** shows the experimental sodium-vacancy ordering as reported by Bianchini et al.[1] in which there are three different sodium positions occupied: the fully occupied Na(1) position [010] and the partially occupied Na(2) and Na(3) position [100] and [110], respectively (see also Figure 1 in the main manuscript). The calculated structures have an energy variation of less than 4 meV/f.u. and always have the Na(1) position occupied. The Na(2) and Na(3) positions are not always occupied, with the Na(2) position being more often occupied than the Na(3) position. Therefore, the experimental structure, which is an average of these calculated structures due to thermal excitations, is accurately reproduced by our calculations.

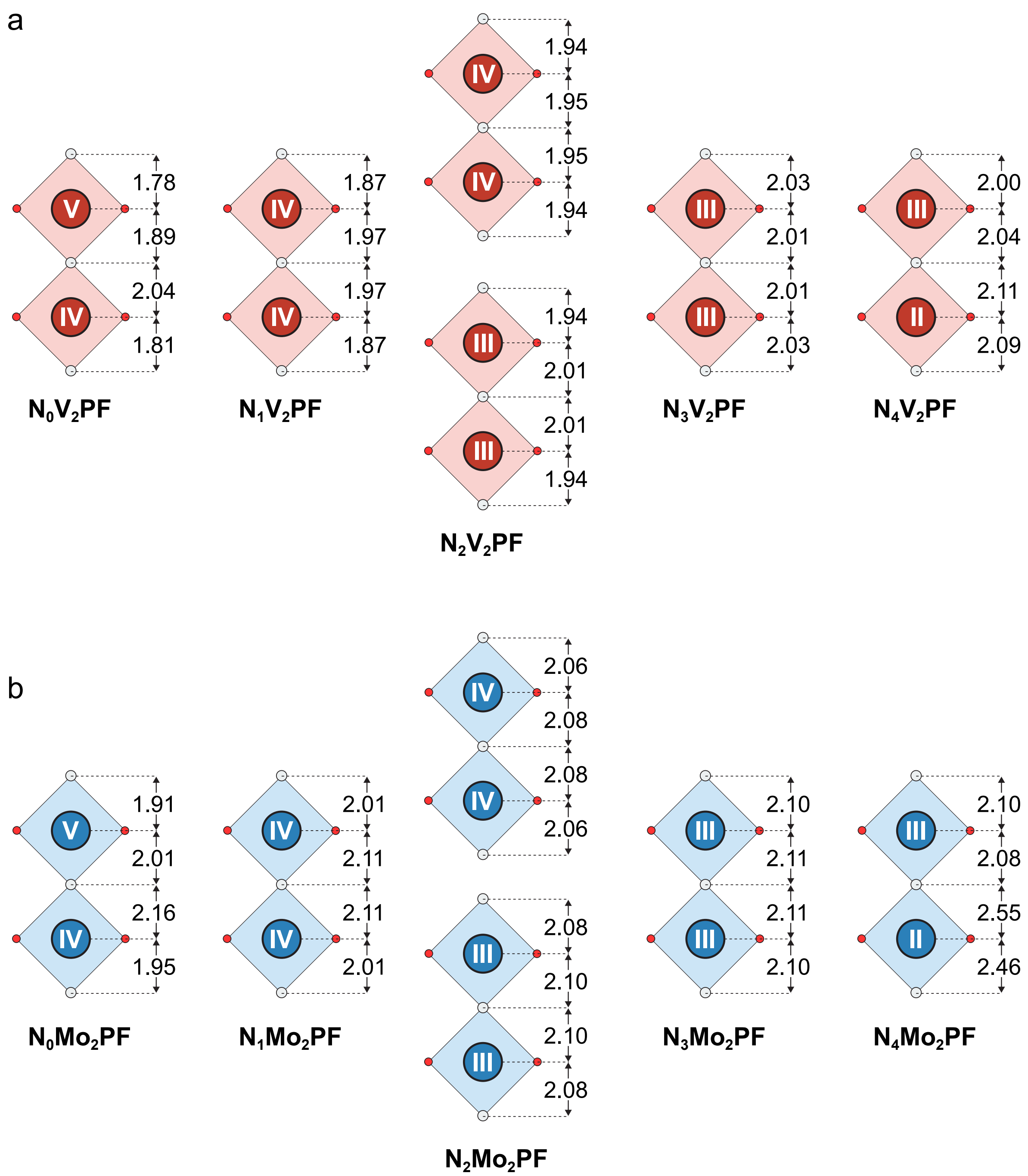

**Supporting Figure S6:** Bioctahedral units for $Na_xV_2(PO_4)_2F_3$ (panel **a**) and $Na_xMo_2(PO_4)_2F_3$ (panel **b**). The oxidation states for the vanadium (shown in red) and molybdenum (shown in blue) are indicated in white. The bond lengths between the fluorine atoms and vanadium (F–V) or molybdenum (F–Mo) atoms are shown in black, measured in Å. Only the most stable structures, as represented on the convex hull in Figure 2 in the manuscript, are visualized. Data obtained at GGA+*U* level of theory.

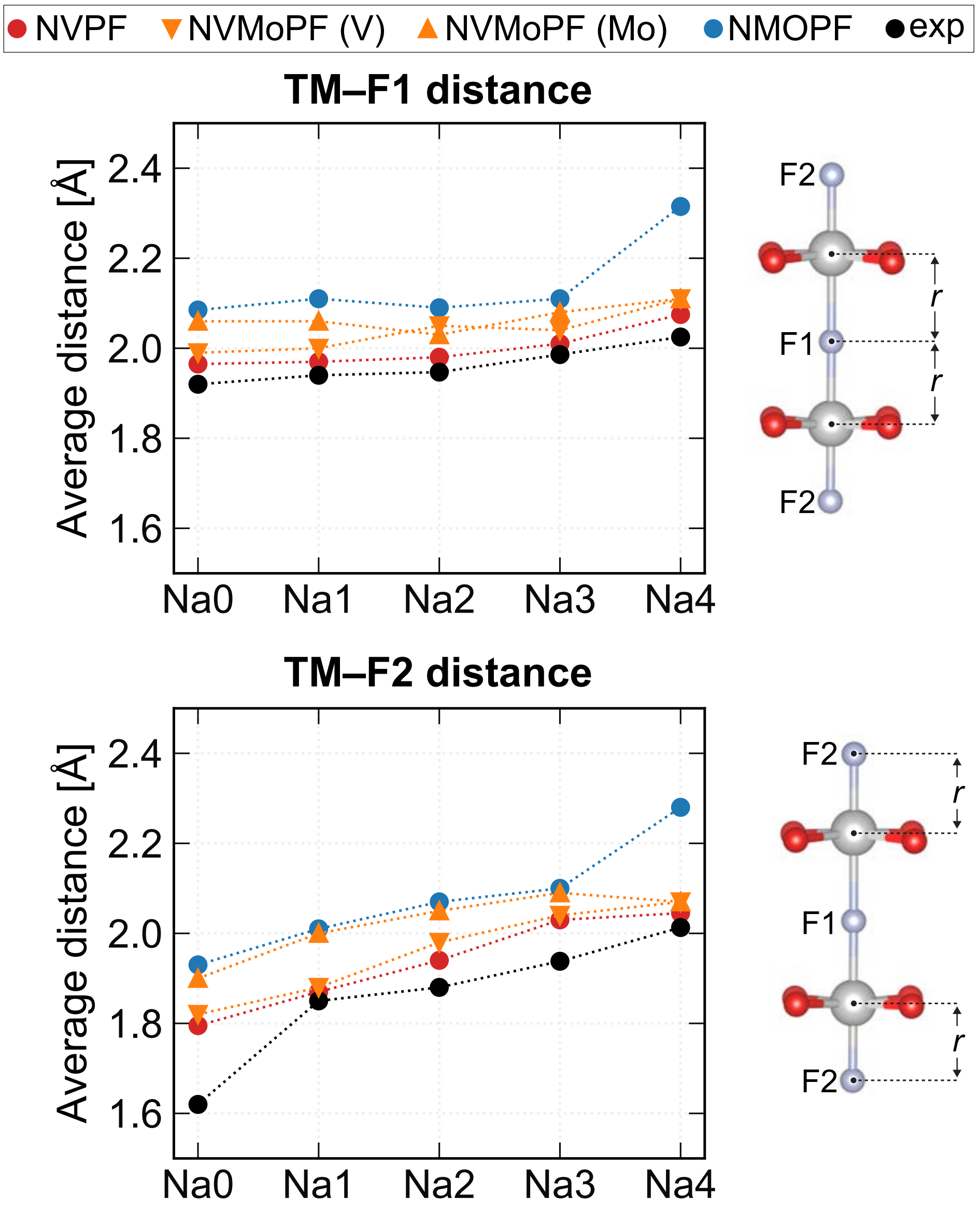

**Figure S7.** Average TM–F1 and TM–F2 distances for NVPF (red), NVMoPF (orange) and NMoPF (blue) obtained in this study at GGA+$U$ level of theory. The experimental values[1–3] are given in black. For NVMoPF, the F–Mo and F–V distances are represented by a triangle pointing up (▲) and triangle pointing down (▼), respectively.

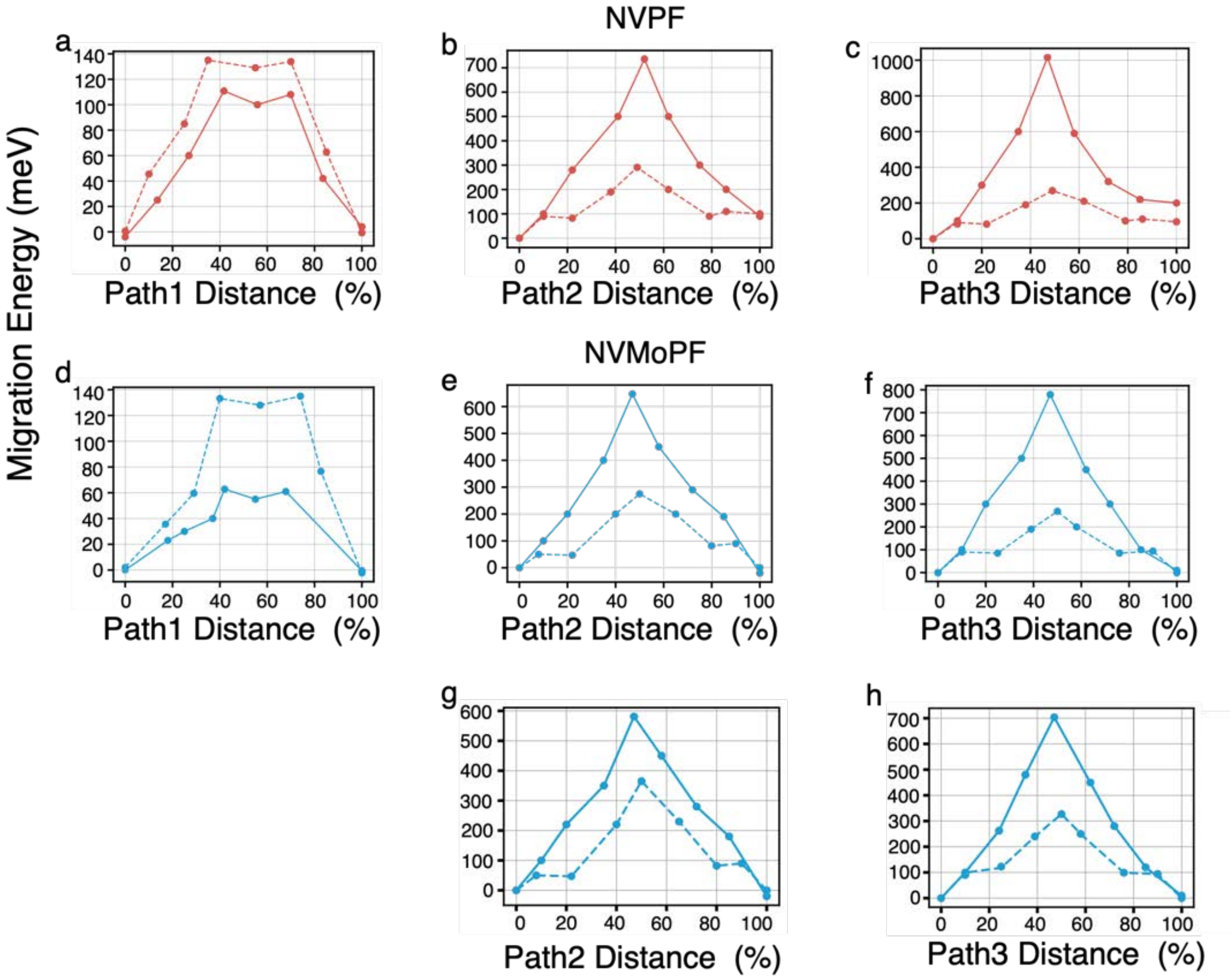

**Supporting Figure S8:** Minimum energy path of NVPF and NVMoPF computed by NEB. Panels **a** and **d** represent the path1, panels b and 3 represent the path2, and panels c and f represent the path3. The results at the sodiated/discharged limit, and desodiated/charged limit are denoted by the solid lines and dashed lines, respectively. In NVMoPF, additional NEB calculations are considered, i.e., the path2 which compasses the $V_2O_8F_3$ bioctahedron (panel **g**), the path2 which compasses the $Mo_2O_8F_3$ bioctahedron (panel **e**); the path3 which connects Na(2)-Na(3)-Na(3)-Na(2) (panel **f**), and the path3 which connects Na(1)-Na(3)-Na(3)-Na(1) (panel **h**).

**Supporting Table S1**. Active redox couples, voltage (V vs. $Na/Na^+$ in Volts), capacity (mAh/g), and energy density (Wh/kg) for NVPF, NVMoPF, NMoPF, NVNbPF, NNbPF, and NWPF, which are extensively discussed in the main article, and the additional $Na_xV_{1.5}Mo_{0.5}(PO_4)F_3$ (25% Mo) and $Na_xV_{0.5}Mo_{1.5}(PO_4)F_3$ (75% Mo) systems. The results are corresponding to each intercalation step from x=0 to x=4. Intercalation voltages were predicted using the GGA+*U* level of theory. The capacities are normalized with the mass of the x=3 composition. The intercalation steps in bold text describe the Na composition region where sodium ions are predicted to be thermodynamically accessible (i.e., the green domains shown in **Figure 2** and **Figure 3** in the main article). The percentages in the energy density column are the relative change in energy density with respect to the experimentally reported energy density for parent NVPF.

| Na content | Redox couple | Voltage | Capacity | Energy density |
|---|---|---|---|---|
| | | **$Na_xV_2(PO_4)_2F_3$ (NVPF)** | | |
| 4 → 3 | $V^{II} \rightarrow V^{III}$ | 1.44 | 64.2 | 92.4 |
| 3 → 2 | $V^{III} \rightarrow V^{IV}$ | 3.24 | 64.2 | 208.0 |
| 2 → 1 | $V^{III} \rightarrow V^{IV}$ | 3.60 | 64.2 | 231.1 |
| 1 → 0 | $V^{IV} \rightarrow V^{V}$ | 4.53 | 64.2 | 290.8 |
| **3 → 1** (*Exp.*[1]) | | **3.95 (avg)** | **128.4** | **507** |
| | | **$Na_xVMo(PO_4)_2F_3$ (NVMoPF)** | | |
| 4 → 3 | $V^{II} \rightarrow V^{III}$ | 1.57 | 57.9 | 90.9 |
| 3 → 2 | $Mo^{III} \rightarrow Mo^{IV}$ | 2.64 | 57.9 | 152.9 |
| 2 → 1 | $V^{III} \rightarrow V^{IV}$ | 3.40 | 57.9 | 196.9 |
| 1 → 0 | $Mo^{IV} \rightarrow Mo^{V}$ | 3.93 | 57.9 | 227.5 |
| **3 → 0** | | **3.31 (avg)** | **173.7** | **577.3 (+13.9%)** |
| | | **$Na_xMo_2(PO_4)_2F_3$ (NMoPF)** | | |
| 4 → 3 | $Mo^{II} \rightarrow Mo^{III}$ | 0.18 | 52.8 | 9.5 |
| 3 → 2 | $Mo^{III} \rightarrow Mo^{IV}$ | 2.61 | 52.8 | 137.8 |
| 2 → 1 | $Mo^{III} \rightarrow Mo^{IV}$ | 3.13 | 52.8 | 165.3 |
| 1 → 0 | $Mo^{IV} \rightarrow Mo^{V}$ | 3.94 | 52.8 | 208.0 |
| **3 → 0** | | **3.23 (avg)** | **158.4** | **511.1 (+0.8%)** |
| | | **$Na_xVNb(PO_4)_2F_3$ (NVNbPF)** | | |
| 4 → 3 | $Nb^{III} \rightarrow Nb^{IV}$ | 1.17 | 58.3 | 68.2 |
| 3 → 2 | $V^{II} \rightarrow V^{III}$ | 1.59 | 58.3 | 92.7 |
| 2 → 1 | $Nb^{IV} \rightarrow Nb^{V}$ | 2.47 | 58.3 | 144.0 |
| 1 → 0 | $V^{III} \rightarrow V^{IV}$ | 3.29 | 58.3 | 191.8 |
| **3 → 0** | | **2.45 (avg)** | **174.9** | **428.5 (–15.5%)** |
| | | **$Na_xNb_2(PO_4)_2F_3$ (NNbPF)** | | |
| 4 → 3 | $Nb^{II} \rightarrow Nb^{III}$ | 0.28 | 53.4 | 15.0 |
| 3 → 2 | $Nb^{III} \rightarrow Nb^{IV}$ | 1.38 | 53.4 | 73.7 |
| 2 → 1 | $Nb^{III} \rightarrow Nb^{IV}$ | 1.75 | 53.4 | 93.5 |
| 1 → 0 | $Nb^{IV} \rightarrow Nb^{V}$ | 2.51 | 53.4 | 134.0 |
| **3 → 0** | | **1.88 (avg)** | **160.2** | **301.2 (–40.6%)** |
| | | **$Na_xW_2(PO_4)_2F_3$ (NWPF)** | | |
| 4 → 3 | – | – | 39.2 | – |
| 3 → 2 | $W^{III} \rightarrow W^{IV}$ | 1.60 | 39.2 | 62.7 |
| 2 → 1 | $W^{III} \rightarrow W^{IV}$ | 2.12 | 39.2 | 83.2 |
| 1 → 0 | $W^{IV} \rightarrow W^{V}$ | 3.06 | 39.2 | 120.0 |
| **3 → 0** | | **2.26 (avg)** | **117.6** | **265.9 (–47.6%)** |
| | | **$Na_xMo_{0.5}V_{1.5}(PO_4)_2F_3$ (25% Mo)** | | |
| 4 → 3 | $V^{II} \rightarrow V^{III}$ | 1.38 | 60.9 | 84.0 |
| 3 → 2.5 | $Mo^{III} \rightarrow Mo^{IV}$ | 2.81 | 30.4 | 85.4 |
| 2.5 → 2 | $V^{III} \rightarrow V^{IV}$ | 3.16 | 30.4 | 96.1 |
| 2 → 1.5 | $Mo^{IV} \rightarrow Mo^{V}$ | 3.62 | 30.4 | 110.0 |
| 1.5 → 1 | $V^{III} \rightarrow V^{IV}$ & $Mo^{IV} \rightarrow Mo^{V}$ | 3.64 | 30.4 | 110.7 |
| 1 → 0 | $Mo^{IV} \rightarrow Mo^{VI}$ | 4.12 | 60.9 | 250.9 |
| **3 → 0** | | **3.47** (avg) | **182.6** | **651.1 (+28.4%)** |
| | **$Na_xMo_{1.5}V_{0.5}(PO_4)_2F_3$ (75% Mo)** | | | |
| 4 → 3 | $V^{II} \rightarrow V^{III}$ & $Mo^{II} \rightarrow Mo^{III}$ | 0.91 | 55.2 | 50.2 |
| 3 → 2 | $Mo^{III} \rightarrow Mo^{IV}$ | 2.50 | 55.2 | 138.0 |
| 2 → 1 | $V^{III} \rightarrow V^{IV}$ & $Mo^{III} \rightarrow Mo^{IV}$ | 3.37 | 55.2 | 186.0 |
| 1 → 0 | $Mo^{IV} \rightarrow Mo^{V}$ | 3.97 | 55.2 | 219.1 |
| **3 → 0** | | **3.28** (avg) | **165.7** | **543.2 (+7.1%)** |

**Supporting Table S2.** Space group (Spg.), lattice parameters (in Å and °) and volumes (in Å$^3$) for $NaV_{2-y}Mo_y(PO_4)_2F_3$ with y=0, 0.5, 1.0, 1.5, and 2.0. All lattice parameters are defined such that a < b << c (in Å), which are not representative of the reported space group.

| **Composition** | **Spg.** | ***a*** | ***b*** | ***c*** | **α** | **β** | **γ** | **V** | ***V/Z*** |
|---|---|---|---|---|---|---|---|---|---|
| | | | | **NVPF (0% Mo)** | | | | | |
| $Na_4V_2(PO_4)_2F_3$ | *Amam* | 9.34 | 9.36 | 10.74 | | | | 939.51 | 226.45 |
| $Na_3V_2(PO_4)_2F_3$ | *Cmc2₁* | 9.15 | 9.18 | 10.88 | | | | 913.97 | 228.37 |
| $Na_2V_2(PO_4)_2F_3$ | *Cmmm* | 9.06 | 9.07 | 10.91 | | | | 896.13 | 224.02 |
| $Na_1V_2(PO_4)_2F_3$ | *Imm2* | 8.93 | 8.94 | 11.23 | | | | 897.10 | 224.2 |
| $V_2(PO_4)_2F_3$ | *I4/mmm* | 8.90 | | 11.28 | | | | 892.24 | 223.06 |
| | | | | **$NV_{1.5}Mo_{0.5}PF$ (25% Mo)** | | | | | |
| $Na_4V_{1.5}Mo_{0.5}(PO_4)_2F_3$ | *Amm2* | 9.42 | 9.43 | 10.78 | | | | 958.95 | 239.74 |
| $Na_3V_{1.5}Mo_{0.5}(PO_4)_2F_3$ | *P1* | 9.22 | 9.26 | 10.91 | 89.9 | 89.9 | 90.04 | 931.54 | 232.89 |
| $Na_2V_{1.5}Mo_{0.5}(PO_4)_2F_3$ | *Amm2* | 9.13 | 9.14 | 11.01 | | | | 917.78 | 229.45 |
| $Na_1V_{1.5}Mo_{0.5}(PO_4)_2F_3$ | *P1* | 9.00 | 9.01 | 11.33 | 90.5 | 90.8 | 91.0 | 459.53 | 229.77 |
| $V_{1.5}Mo_{0.5}(PO_4)_2F_3$ | *Amm2* | 8.96 | 8.97 | 11.42 | | | | 917.83 | 229.46 |
| | | | | **NVMoPF (50% Mo)** | | | | | |
| $Na_4VMo(PO_4)_2F_3$ | *Cmmm* | 9.50 | 9.52 | 10.82 | | | | 978.76 | 244.70 |
| $Na_3VMo(PO_4)_2F_3$ | *Cm* | 9.28 | 9.32 | 10.97 | | | 90.07 | 948.95 | 237.24 |
| $Na_2VMo(PO_4)_2F_3$ | *Cmmm* | 9.18 | 9.18 | 11.12 | | | | 936.43 | 231.11 |
| $Na_1VMo(PO_4)_2F_3$ | *Cm* | 9.07 | 9.08 | 11.44 | | | 90.8 | 941.33 | 235.33 |
| $VMo(PO_4)_2F_3$ | *Cmmm* | 9.03 | 9.04 | 11.54 | | | | 941.50 | 235.38 |
| | | | | **$NV_{0.5}Mo_{1.5}PF$ (75% Mo)** | | | | | |
| $Na_4V_{0.5}Mo_{1.5}(PO_4)_2F_3$ | *Pm* | 9.37 | 9.76 | 11.02 | | 93.97 | | 1005.38 | 251.35 |
| $Na_3V_{0.5}Mo_{1.5}(PO_4)_2F_3$ | *P1* | 9.36 | 9.49 | 10.99 | 89.9 | 89.96 | 89.96 | `966.76 | 241.69 |
| $Na_2V_{0.5}Mo_{1.5}(PO_4)_2F_3$ | *P1* | 9.25 | 9.28 | 11.22 | 89.8 | 89.8 | 90.01 | 963.62 | 240.91 |
| $Na_1V_{0.5}Mo_{1.5}(PO_4)_2F_3$ | *P1* | 9.14 | 9.15 | 11.51 | 89.9 | 90.5 | 90.3 | 962.24 | 240.61 |
| $V_{0.5}Mo_{1.5}(PO_4)_2F_3$ | *Amm2* | 9.11 | 9.13 | 11.63 | | | | 967.83 | 241.96 |
| | | | | **NMoPF (100% Mo)** | | | | | |
| $Na_4Mo_2(PO_4)_2F_3$ | *P2₁/m* | 9.34 | 9.83 | 11.35 | | 90.01 | | 1038.02 | 259.51 |
| $Na_3Mo_2(PO_4)_2F_3$ | *Cmc2₁* | 9.44 | 9.44 | 11.05 | | | | 989.70 | 247.43 |
| $Na_2Mo_2(PO_4)_2F_3$ | *Cm* | 9.31 | 9.32 | 11.34 | | 90.03 | | 984.79 | 246.29 |
| $Na_1Mo_2(PO_4)_2F_3$ | *Cmc2₁* | 9.21 | 9.22 | 11.61 | | | | 986.14 | 246.54 |
| $Mo_2(PO_4)_2F_3$ | *I4/mmm* | 9.16 | | 11.84 | | | | 993.66 | 248.42 |